\documentclass[12pt]{article}
\usepackage{amsbsy,graphicx,latexsym}
\newcommand{\beq}{\begin{equation}}
\newcommand{\eeq}{\end{equation}}
\newcommand{\bea}{\begin{eqnarray*}}
\newcommand{\eea}{\end{eqnarray*}}
\newcommand{\beaq}{\begin{eqnarray}}
\newcommand{\eeaq}{\end{eqnarray}}
\catcode`\@=11 \@addtoreset{equation}{section}

\catcode`\@=11
\def\section{\@startsection{section}{1}{\z@}{3.5ex plus 1ex minus
   .2ex}{2.3ex plus .2ex}{\large\bf}}

\begin{document}
\begin{flushright} {KIAS-P03095}\end{flushright}
\begin{flushright}{hep-th/0312308} \end{flushright}

\vspace{10mm}
\centerline{\Large \bf Perturbation theory of the space-time }
\centerline{\Large \bf non-commutative real scalar field theories}
\vskip 1cm
\centerline{\large
Chaiho Rim $^{1}$, Yunseok Seo$^{2}$
and Jae Hyung Yee$^{3}$ }
\vskip 1cm\centerline{\it $^{1}$ Department of Physics, 
Chonbuk National University}
\centerline{\it Chonju 561-756, Korea}
\centerline{\it rim@mail.chonbuk.ac.kr}
\vskip .5cm
\centerline{\it $^{2}$ Department of Physics, 
Kyung Hee University}
\centerline{\it Seoul 130-701, Korea}
\centerline{\it seo@khu.ac.kr}
\vskip .5cm
\centerline{\it $^{3}$ Institute of Physics and Applied Physics, 
Yonsei University}
\centerline{\it Seoul 120-749, Korea}
\centerline{\it jhyee@phya.yonsei.ac.kr}
\vskip 2cm
\centerline{\bf Abstract}
\vskip 0.5cm
\noindent
The perturbative framework of the space-time non-commutative
real scalar field theory is formulated,
based on the unitary S-matrix.
Unitarity of the S-matrix is explicitly checked order by order 
using the Heisenberg picture of 
Lagrangian formalism of the second quantized operators, 
with the emphasis of the so-called minimal realization 
of the time-ordering step function
and of the importance of the $\star$-time ordering.
The Feynman rule is established and is presented using 
$\phi^4$ scalar field theory. It is shown that 
the divergence structure of space-time 
non-commutative theory is the same as the one 
of space-space non-commutative theory, 
while there is no UV-IR mixing problem in this 
space-time non-commutative theory.

\newpage

\section{Introduction}
\noindent

Non-commutative field theory (NCFT) \cite{szabo} 
is the field theory defined 
on the {\sl non-commutative} (NC) coordinates. 
We will consider NC coordinates which obey
\beq
[x^\mu, x^\nu] = i \theta^{\mu\, \nu}
\label{xmunu}
\eeq
where $\theta^{\mu\nu} $ is an antisymmetric c-number.
Space non-commutative theory (SSNC) involves
only the space non-commuting coordinates
( $\theta^{0\nu} =0$),
whereas space-time non-commutative theory (STNC)
contains the non-commuting time 
( $\theta^{01} \ne 0$).

The non-commuting nature of the coordinates
is naturally adapted to the operator formalism 
for the first quantization of the theory.
However, the operator formalism is not convenient 
for second quantized field theory.
Fortunately, there exists another formalism
suited for NCFT based on the Weyl's idea \cite{weyl}:
NCFT is constructed using the $\star$-product of fields
but space-time coordinates are {\sl commuting}.
The $\star$-product encodes all the non-commuting
nature of the theory and fixes the ordering ambiguity 
of non-commuting coordinates.
We adopt the Moyal product \cite{moyal}
as the $\star$-product representations,
\beq
f\star g\, (x)=
e^{\frac{i}{2}{\partial}_x \wedge
{\partial}_y}f(x)g(y)|_{y=x} \,
\eeq
with $a \wedge b \equiv \theta^{\mu\nu} a_\mu  b_\nu$.
In terms of the $\star$-product, the non-commuting
nature of the coordinates in (\ref{xmunu})
is written as
\bea
[x^\mu \stackrel{\star}, x^\nu]
\equiv
x^\mu \star x^\nu - x^\nu \star x^\mu
= i \theta^{\mu\, \nu} \,
\eea
and now the coordinates itself $x^\mu$ is commuting each  other.
The merit of the Moyal product is that it does not change
the kinetic term of the action even after
introducing the $\star$-product,
and allows the conventional perturbation
with respect to the free theory \cite{ncperturb}\,.

NCFT is a non-local field theory and the theory
behaves very differently in many respects.
Lorentz symmetry is usually broken
even though there are some attempt to cure this\cite{dfr}.
Especially, STNC is known to have
micro-causality problem \cite{sst}
and unitarity problem \cite{j-m}
due to the infinite number of
time-derivatives.

Among these problems, it is proposed
in \cite{bahns,ry,topt}
that the unitarity problem can be avoided
if one uses the careful time-ordering in the S-matrix.
However, each proposal has different aspects,
which needs to be distinguished from each other.
The proposal by \cite{bahns} is the first attempt to solve
the unitarity problem and pointed out
the unitarity problem is not inherent to the theory
but due to the formalism of the theory.
The proposal provides the lowest order S-matrix,
which needs higher order derivative correction
to make the proper S-matrix.
After this correction, however,
it turns out \cite{ry} 
that the time-ordering should be done
before the $\star$-operation 
in contrast to their proposal.
The proposal by  \cite{topt} is 
called the time-ordered perturbation theory 
but is pointed out in \cite{nonunitary}
that the gauge invariance may not be
respected when applied to a gauge theory.

Our proposal in \cite{ry} is critically different
from the other two in the sense of the time-ordering.
The time-ordering is done 
in terms of the so-called minimal realization
and the time-ordering should be performed
before the $\star$-operation.
The purpose of this paper is to clarify and justify the
time-ordering in the S-matrix of
STNC QFT proposed in \cite{ry} 
and is to construct the systematic 
formalism of the perturbation theory.

In section \ref{s}, S-matrix is explicitly constructed
using the Lagrangian of the second quantized 
operator in the Heisenberg picture 
following Yang and Feldman \cite{yf}.
Even though the unitary transformation at 
{\it finite time} could not be found, there exists
the unitary S-matrix.
The unitarity of the S-matrix is explicitly proven
order by order in the coupling constant.
In section \ref{f},
Feynman rule is established and
perturbation theory is formulated
using the real scalar $\phi^4$ theory.
Section \ref{c} is the conclusion and outlook.

\section{S-matrix for scalar STNC field theory \label{s}}

\subsection{$\star$-operation and interaction Lagrangian}
\noindent

The Lagrangian of a real scalar STNC field theory
consists of the free part and interacting part.
Using the starred notation,
\bea
\phi_\star^p = \phi \star \phi \star \cdots \star \phi\,
\eea
the interaction Lagrangian in $D-1$ dimensional 
space is given as
\beq
L_I (t) = \int d^{D-1}x \,\,
{\cal L}_I (\phi_\star (x))\,,\qquad
{\cal L}_I (\phi_\star (x))
= -\frac{g}{p!} \,
 \phi_\star^p (x)
\label{p-lagrangian}
\eeq
where $g$ is a coupling constant
and the integration is done over 
the $D-1$ space dimension.

It is convenient to introduce a $\star$-operator
${\cal F}_x$:
\beq
{\cal F}_x
\Big(\phi^p (x) \Big)
\equiv  \phi(x)  \stackrel{x} \star \phi (x) \stackrel{x} \star
\cdots \stackrel{x} \star \phi(x)
= \phi_\star^p (x) \,.
\eeq
The notation $ \stackrel{x} \star$ is to put down  the explicit  argument
which is to be starred. The interaction Lagrangian
density 
$ {\cal L}_I (\phi_\star (x)) $ 
is related through the $ {\cal F}_x $
with an un-starred interaction density
${\cal V} (\phi (x)) $ :
\beq
{\cal L}_I (\phi_\star (x)) 
= {\cal F}_x \Bigg ({\cal V} \Big(\phi (x)\Big) \Bigg)
\eeq
where
\beq
{\cal V} (\phi (x))
\equiv -\frac{g}{p!} \,\, \phi^{p} (x) \,.
\eeq

It is worth to mention
that the un-starred quantity uniquely
defines the starred quantity
\bea
A(x)\star B(x) = {\cal F}_x \Big( A(x) B(x) \Big) \,.
\eea
However, the inverse is not true since, even if
$ A(x) B(x) = B(x) A(x)$, the starred one is not;
\beq
{\cal F}_x \Big ( A(x) B(x) \Big)
\ne  {\cal F}_x \Big ( B(x) A(x) \Big) \,,
\label{star-ordering}
\eeq
because of the non-commuting nature of the star-product.
This ordering ambiguity is to be treated carefully.
A composite $\star$-operator can be also defined
\beq
{\cal F}_{xy} \equiv {\cal F}_x {\cal F}_y \,,
\eeq
which is commutative
\beq
{\cal F}_{xy}  = {\cal F}_{yx}\,.
\eeq

\subsection{Out-field}

The field at an arbitrary time
is obtained from the field equation
\beq
(\Box + m^2 ) \,\phi (x) = \xi_\star( \phi(x))
\label{phi-eq}
\eeq
where $\xi_\star$ is the functional of fields,
derived from the interaction Lagrangian (\ref{p-lagrangian}),
\beaq
\xi_\star (\phi(x))
&\equiv&
\frac{\delta}{\delta \phi(x)} \int dt L_I (t)
\nonumber\\
&=& - \frac{g}{p!}\, \int d^D y\,
\Bigg( \delta(x-y) \stackrel{y}\star \phi_\star^{p-1}(y)
+
\phi(y) \stackrel{y}\star \delta(x-y)
\stackrel{y}\star \phi_\star^{p-2}(y)
\nonumber\\
&&\qquad + \cdots +
\phi_\star^{p-2}(y) \stackrel{y}\star \delta(x-y)
 \stackrel{y}\star \phi(y) +
\phi_\star^{p-1}(y)\stackrel{y}\star \delta(x-y) \Bigg) \,.
\eeaq
Introducing a compact notation for the 
symmetrization of $n$ distinctive quantities,
\bea
\Big\{ \prod_{i=1}^n  A_i \Big\}_s
\equiv  \sum_{\rm s(1,2, \cdots, n)}
\prod_{i=1}^n  A_{s(i)} \,,
\eea
where $s(1,2,\cdots, n)$
is the permutation of $1,2, \cdots,n$,
we may put $\xi_\star $ as
\beq
\xi_\star(\phi(x))
= - \frac{g}{p!}\, \int d^D y\, {\cal F}_y
\Big\{ \delta(x-y) \phi^{p-1} (y) \Big\}_{s(y)}
\eeq
where the subscript $s(y)$ refers to the symmetrization of
operators with argument $y$.

The solution of Eq.~(\ref{phi-eq}) is given as
\beaq
\phi (x)
&=& \phi_{\rm in}(x) +
\int d^D y \, \bigtriangleup_{\rm ret} (x-y) \,
\xi_\star (\phi(y))
\nonumber\\
&=&\phi_{\rm out}(x) +
\int d^D y \, \bigtriangleup_{\rm adv} (x-y) \,
\xi_\star (\phi(y))\,.
\label{phi} 
\eeaq
Here $\Delta_{\rm ret} (x) $
($\Delta_{\rm adv} (x) $) denotes
the retarded (advanced) Green's function,
\beq
\Delta_{\rm ret}  (x) 
= -\theta(x^0 )\, \Delta (x)\,,
\qquad
\Delta_{\rm adv} (x) = \theta(-x^0 )\,\Delta (x)
\label{ad-ret}
\eeq
and $\Delta (x)$ is the free commutator function,
\beq
[\phi_0 (x), \phi_0 (0) ] = i \Delta (x)\,.
\label{freecommutator}
\eeq
Employing the delta-function identity,
\beq
\int d^D y \, \int d^D z\, A(x-y)
\Big( \delta (y-z) \stackrel{z}\star B (z) \Big)
=  \int d^D y \, \Big( A(x-y) \stackrel{y}\star B (y)\Big)
\eeq
for arbitrary function of $A(x-y)$ and an operator $B(y)$,
we may put the retarded or advanced Green's function
of (\ref{phi}) inside the star-operation:
\beaq
\phi (x)
&=& \phi_{\rm in}(x) +
\int d^D y \, {\cal F}_y
\Bigg\{ -\frac g{p!}
\Delta_{\rm ret} (x-y) \,\phi^{p-1}(y)
\Bigg\}_{s(y)}
\nonumber\\
&=& \phi_{\rm out}(x) +
\int d^D y \, {\cal F}_y
\Bigg\{ -\frac g{p!}
\Delta_{\rm adv} (x-y) \, \phi^{p-1}(y)
\Bigg\}_{s(y)} \,.
\label{phi-in-out}
\eeaq

This gives the relation between the out-field
and the in-field,
\beq
\phi_{\rm out}(x)
= \phi_{\rm in}(x)
+ \int d^D y \, {\cal F}_y
\Bigg\{ \frac g{p!}
\Delta (x-y) \phi^{p-1}(y) \Bigg\}_{s(y)} \,.
\eeq
Therefore, the out-field is written iteratively
in terms of the in-field if one uses the
relation in(\ref{phi-in-out}):
Putting $\phi$ as
$ \phi  = \phi_0 + \phi_1 + \phi_2 \cdots $
where $\phi_n$
represents the order of $g^n$ contribution.
A few explicit forms of $\phi_n$'s are given as
\beaq
\phi_0 (x) &=& \phi_{\rm in} (x)
\nonumber\\
\phi_1 (x) &=& \int d^D y \, \Delta_{\rm ret} (x-y)\,
\xi_\star (\phi_0(y))
\cr  
&=& \int d^D y \,
{\cal F}_y \Bigg \{
 -\frac{g}{ p! } \,
\Delta_{\rm ret} (x-y)\,\phi_0^{p-1}(y)\Bigg\}_{s(y)}
\nonumber\\
\phi_2 (x)
&=&\int d^D y \,  {\cal F}_y
\Bigg\{  -\frac{g}{p!}  \Delta_{\rm ret} (x-y) \,
\phi_0^{p-2}(y) \phi_1 (y)\Bigg\}_{s(y)}
\nonumber\\
\phi_3 (x)  &=& \int d^D y \, \, {\cal F}_y \Bigg\{
-\frac{g}{p!} \,
\Delta_{\rm ret} (x-y) \Big( \phi_0^{p-2}(y) \phi_2 (y)
+ \phi_0^{p-3}(y) \phi_1^2 (y)\Big ) \Bigg\}_{s(y)}
\nonumber\\
\phi_n (x)  &=& \int d^D y\,
\, {\cal F}_y \Bigg\{
\sum_{q_1 +\,\cdots +q_{p-1}=n-1} \,
-\frac g{p!} \, \Delta_{\rm ret} (x-y) \, \phi_{q_1}(y)
\cdots \phi_{q_{p-1}}(y)  \Bigg\}_{s(y)} \,.
\label{phiout}
\eeaq
For later use, we put the explicit form
of the out-field as
\bea 
\phi_{\rm out}
= \sum_{i=0}^\infty \varphi_{i} (x) \,,
\eea
where $\varphi_0= \phi_0$
and for $n\ge 1$
\beq
\varphi_{n} (x)
= \int d^D y\,
\, {\cal F}_y \Bigg\{
\sum_{q_1+\,\cdots +q_{p-1}=n-1} \,
\frac g{p!} \, \Delta (x-y)  \phi_{q_1}(y)
\cdots \phi_{q_{p-1}} (y) \Bigg\}_{s(y)} \,.
\label{varphiout}
\eeq

\subsection{S-matrix}

The S-matrix relates the out-field with the in-field:
\beq
\phi_{\rm out} = S^\dagger \,\phi_{\rm in}\,S  \,.
\eeq
With the notation $S = e^{i\delta}$,
the out-field would be written as
\beq
\phi_{\rm out}
= \phi_{\rm in} + [\,\phi_{\rm in}, i \delta \,]
+ \frac12 [[\,\phi_{\rm in} , i \delta\, ], i\delta\, ]
+ \cdots \,.
\label{S-delta}
\eeq
The first order term in $g$ should be written as
\beq
[\, \phi_{\rm in}, i \delta \,]
=\varphi_1 (x)
= \int d^D y\,
\, {\cal F}_y \Bigg\{ \frac g{p!}
\Delta (x-y)  \phi_0^{p-1} (y)
\Bigg\}_{s(y)} 
\eeq
and determines the phase $\delta$ to the first order in $g$,
\beq
\delta = \int d^D y\,
\, {\cal F}_y \Bigg( - \frac g{p!}
\phi_0^{p} (y) \Bigg)
+ O(g^2)
=  \int d^D y\,
 {\cal L}_I \Big(\phi_{0\star} (y)\Big)
+ O(g^2)\,.
\eeq

Higher order solutions require the time-ordering
as in the ordinary field theory.
However, the time-ordering needs a special care 
in the $\star$-product and a consistent unitary S-matrix
is proposed in  \cite{ry} as 
\beq
S= \sum_{n=0}^\infty i^n  A_n
\label{S-matrix}
\eeq
where $A_n$ is the order of $g^n$ with $A_0 =1$:
\beq
A_n = \int\!\!\cdots\!\!\int
d^D x_1\! \cdots \! d^D x_n \,\,
{\cal F}_{1 \cdots n}
\Bigg(
\theta_{12\cdots n}\,\,\,
{\cal V} \Big(\phi_0(x_1) \Big) \cdots {\cal V} 
\Big(\phi_0(x_n)\Big) \Bigg)
\label{An}
\eeq
where we use the composite version of $\star$-operation
\bea
{\cal F}_{1 2\cdots n}
\equiv {\cal F}_{x_1}  {\cal F}_{x_2} \cdots  {\cal F}_{x_n}\,,
\eea
whose operation is independent of
the permutation of operators.

The time-ordering is given in terms of the step function,
\bea
\theta_{12\cdots n} \equiv \theta (t_1 -t_2) \,
\theta (t_2- t_3) \cdots \theta(t_{n-1} -t_n)\,.
\eea
The ambiguity of the time-ordering 
is due to the point splitting ambiguity
of the arguments in $\theta(x^0 -y^0)$. 
For example,
one might have 
\beq
{\cal F}_{x y}
\Bigg(\, \theta(x^0-y^0)\, \phi^p (x)\, \phi^q (y)
\Bigg)  \ne
{\cal F}_{x y}
\Bigg(\, \theta(x^0-y^0)\, \phi^p (x)\,
\theta(x^0-y^0)\, \phi^q (y) \Bigg) \,,
\eeq
depending on how one splits the coordinates 
to define the proper $\star$-product. 
We fix this ambiguity by using the so-called
{\it minimal realization}
of the step-function  in the $\star$-operation:
The minimal realization of $\star$-operation
is to change the step function
$\theta(x^0 -y^0)$ to $ \theta(x_i^0 -y_j^0)$
and is to use the step function {\it only once};
\beaq
&&{\cal F}_{x y}
\Big(\, \theta(x^0-y^0)\, \phi^p (x) \,
\phi^q (y)  \,\Big)
\nonumber \\
&& \quad =
{\cal F}_{x y}
\Bigg(\, \theta(x_i^0-y_j^0) \,
\phi(x_1) \! \cdots \! \phi(x_i)
\! \cdots \! \phi (x_p)
\phi(y_1)\! \cdots \!\phi(y_j)\!
\cdots \!  \phi (y_q)
\Big|_{x_i = x,\, y_j=y}\Bigg) \,.
\eeaq
The split coordinates of $\theta(x^0 -y^0)$
is to be assigned {\it a posteriori}
as the argument of the spectral function
$ \Delta(x_i^0 -y_j^0)$ 
which connects two vertices.
And even in the presence of many spectral functions
we have only one step function,
\beq
\theta(x^0 -y^0) \prod_{a,b} \Delta(x_a -y_b)
\, \longrightarrow \,
\theta(x_i^0 -y_j^0) \prod_{a,b}  \Delta(x_a -y_b)
\eeq
where $i\, (j)$ is just one of indices among $a$'s ($b$'s).
The minimal realization assumption sounds {\it ad hoc},
but is necessary to prove the relation
$\phi_{\rm out} = S^\dagger \phi_{\rm in} S$
in section \ref{s-in-out}.
This minimal realization of the step-function 
is the crucial difference from the 
recipe given in the time-ordered perturbation 
theory given in \cite{topt}.

Introducing $\star$-time-ordering $T_{\star}$ as
\beq
T_{\star} \Big\{A(t_1) A(t_2) \Big\} =
{\cal F}_{12} \Bigg( \theta_{12} \, A(t_1) A (t_2)
+ \theta_{21}
\, A(t_2) A (t_1)\Bigg)\,.
\eeq
we may put the S-matrix in a compact form as
\beaq
S &=& \sum_{n=0}^\infty \, \frac{i^n}{n!} \,
\int d^D x_1
\cdots \int d^D x_n \,
T_{\star} \Bigg (\, 
{\cal V} \Big(\phi_0 (x_1)\Big)
\cdots  {\cal V} \Big(\phi_0 (t_n)\Big) \Bigg)
\nonumber\\ &\equiv&
T_{\star}
\exp \Bigg( i\int d^Dx \,
{\cal V}\Big(\phi_0 (x) \Big) \Bigg)\,.
\eeaq

\vskip 0.5cm
\subsection{S-matrix and in- and out-field 
\label{s-in-out}}
\noindent

In this section, we check that
$S^\dagger \phi_{\rm in}(x)  S $
reproduces the correct out-field given in  
(\ref{varphiout}).
For this purpose,
we evaluate the out field
using the S-matrix definition (\ref{S-matrix})
and denote it as $ \Phi_{\rm out}(x) $:
\beq
\Phi_{\rm out}(x) \equiv
S^\dagger \phi_{\rm in}(x)  S
= \sum_{n=0}^\infty \Phi_{(n)} (x)
\eeq
where $\Phi_{(n)} $ is the out-field term
of order $g^n$. Each order is given as
\beq
\Phi_{(0)} (x)  =\phi_0 (x) \,,\qquad
\Phi_{(n)} (x) =
\sum_{\ell, m \atop \ell+m=n} 
(-)^\ell (i)^{\ell+m } \,
A_\ell^\dagger \,\,\phi_0(x)\,\, A_m  \,.
\eeq
This result is to be compared with the out-field
obtained from the equation of motion, $\varphi_n$ in
(\ref{varphiout}).
The evaluation at the order of $g$ is given as
\beaq
\Phi_{(1)} 
&=& i \,\Big( \phi_0(x) \,A_1 
- A_1^{\dagger}\, \phi_0(x)\Big)
\cr 
&=&  i\int d^Dy \,
{\cal F}_y \Big(\,  [\,\phi_0(x), 
{\cal V} \Big(\phi_0(y)\Big)\, ] \,\Big)
\cr 
&=& - \Big(- \frac g{p!}\Big) \int d^D y \,
{\cal F}_y  \Bigg(
\Big\{\Delta(x-y) \,\phi_0^{p-1} (y) \Big\}_{s(y)} \Bigg)
\cr 
&=& - \int d^D y \, \Delta (x-y)
\xi_\star ( \phi_0(y)) =\varphi_1 \,.
\eeaq
At the order of $g^2$, the out-field is given as
\beaq
\Phi_{(2)} &=&
i^2 \,\Big(\phi_0(x) \,A_2 
- A_1^\dagger\, \phi_0 A_1 
+ A_2^\dagger\, \phi_0(x) \Big)
\cr 
&=&  i^2 \int d^D y_1 d^D y_2 \,
{\cal F}_{12}\,
\Bigg(  \theta_{12} \, \phi_0(x)\,
{\cal V}(\phi_0 (y_1)) {\cal V} (\phi_0(y_2))
\cr 
&& \qquad\qquad
- {\cal V}_1(\phi_0(y_1))\phi_0 {\cal V}_1(\phi_0(y_2))
+ \theta_{12} \, 
{\cal V}(\phi_0(y_2)) {\cal V}(\phi_0(y_1)) \phi_0(x)
\Bigg)
\cr 
&=&  i^2 \int d^D y_1 d^D y_2 \, {\cal F}_{12}\,
\Bigg(  \theta_{12} \, \Big[[\,\phi_0(x)\,, 
{\cal V}(\phi_0(y_1))\,]\,,
{\cal V}(\phi_0(y_2))\,\Big] \Bigg) \,.
\eeaq
We note that all the quantum operators inside the
$\star$-operation are properly ordered except the
time-ordering step function.
This ordering ambiguity of the step function
will be settled using the
minimal realization.

First note that the commutator
$ \Big[[\,\phi_0(x)\,, \phi_0^p(y_1)\,],
\phi_0^p(y_2)\,\Big]$
is given as
\beaq
\Big[[\,\phi_0(x)\,, \phi_0^p(y_1)\,],
\phi_0^p(y_2)\,\Big]
&=&  i \Big[[\,\{\Delta(x-y_1)\,
\phi_0^{p-1} (y_1) \}_{s1}\,,
\phi_0^p(y_2) \Big]
\nonumber \\&=&
i \Big\{ \Delta(x-y_1) \,\phi_0^{p-2} (y_1)\,
[\phi_0(y_1)\,, \phi_0^p(y_2)] \Big\}_{s1}
\nonumber \\&=&
- \Bigg\{\Delta(x-y_1)\, \phi_0^{p-2} (y_1)\,
\Big\{\Delta(y_1-y_2)\, \phi_0^{p-1}(y_2)
\Big\}_{s2} \Bigg\}_{s1} \,.\qquad\qquad
\eeaq
$s1$ ($s2$) refers to the symmetrization with respect to
$y_1$ ( $y_2$) fields and functions.
Therefore, at each step there is no position ambiguity
for this operators and functions.
 Next, the presence of the time-ordering step function
$\theta_{12}$ changes the commutator function
into the retarded Green's function,
\beq
\theta_{12}\, [[\phi_0(x)\,, \phi_0^p(y_1)\,], \phi_0^p(y_2)\,]
= \Bigg\{\Delta(x-y_1) \phi_0^{p-2} (y_1)
\Big\{\Delta_{\rm ret} (y_1-y_2)
\phi_0^{p-1}(y_2) \Big\}_{s2} \Bigg\}_{s1} \,.
\eeq
Here we used the minimal realization
since we put the step function as the specific position
corresponding to the spectral function, $\Delta (y_1-y_2)$.

Finally, the $\star$-operation on the $s2$ symmetrized part
will give  $\phi_1(y_1) $ in (\ref{phiout}) :
\beaq
&& \int d^D y_2 \, {\cal F}_{2} \,
\Bigg( \theta_{12}
\Big[ [\, \phi_0(x)\,, \phi_0^p(y_1)\, ]\,,
{\cal V} (\phi_0(y_2))\, \Big]\Bigg)
\cr 
&&\qquad  = \Big(-\frac g{p!}\Big)
\Bigg\{\Delta(x-y_1)\, \phi_0^{p-2} (y_1) \,
\int d^D y_2\,  {\cal F}_2 \,
\Big( \{\Delta_{\rm ret} (y_1-y_2) \,\phi_0^{p-1}(y_2) \}_{s2} \Big)
\Bigg\}_{s1}
\cr 
&&\qquad = 
\Big\{\,\Delta(x-y_1) \,\phi_0^{p-2} (y_1)
\, \phi_1 (y_1) \,\Big\}_{s1} \,. 
\eeaq
Therefore, the out-field $\Phi_{(2)} (x) $ reduces 
to $\varphi_{(2)} (x)$ in (\ref{varphiout});
\beq
\Phi_{(2)} (x) =
- \Big(-\frac g{p!} \Big) \,\int dy\,
{\cal F}_y \Bigg( \Delta(x-y) \,\Big \{
\phi_{0}^{p-2}(y)\, \phi_1(y) \Big\}_s  \Bigg)
= \varphi_2 (x)\,.  
\eeq

At the order of $g^3$ the out-field is given as
\beaq
\Phi_{(3)}(x) &=&
i^3\, 
\Big(\phi_0(x)\, A_3 
- A_1 ^\dagger\, \phi_0 (x)\, A_2
+ A_2 ^\dagger\, \phi_0(x)\, A_1
- A_3 ^\dagger\, \phi_0 (x)
\Big)
\cr 
&=&  i^3 \int\!\!\int\!\!\int
d^D y_1 \,d^D y_2 \,d^D y_3  \,\,
\cr 
&&\,\, \times {\cal F}_{123}
\Bigg(  \theta_{123} \, \phi_0(x)\, 
{\cal V}(y_1)\, {\cal V}(y_2)\, {\cal  V}(y_3)
- \theta_{23} {\cal V}(y_1)\, \phi_0(x)\, 
{\cal V}(y_2) \, {\cal V}(y_3)
\cr 
&&\quad 
+ \theta_{12} \, {\cal V}(y_2)\, {\cal V}(y_1) \,
\phi_0(x)\, {\cal V}(y_3)
- \theta_{123}\, {\cal V}(y_3)\,{\cal V}(y_2)\, 
{\cal V}(y_1)\,\phi_0(x) \Bigg)
\cr  
&=& i^3 \int\!\!\int\!\!\int
d^D y_1 d^D y_2 d^D y_3  
\,\, {\cal F}_{123}\,  \Bigg( \theta_{123} \,\, 
[[[\phi_0(x) , {\cal V}(y_1)]\,, {\cal V}(y_2)]\,, 
{\cal V} (y_3)]
\Bigg)\,. \qquad\quad 
\eeaq

Evaluation of the commutator 
$[[[\phi_0(x)\, , {\cal V}(y_1)]\,, 
{\cal V}(y_2)]\,, {\cal V}(y_3)]$
is done in a few steps:
\beaq 
&&
\Bigg[\Big[[\phi_0(x)\, , 
\phi_0^p(y_1)]\,, \phi_0^p(y_2)\Big]\,, 
\phi_0^p(y_3)\,\Bigg]
\nonumber\\&&\quad
= i \Bigg[ \Big[\{ \Delta(x-y_1)\, \phi_0(y_1)^{p-1}\}_{s1},
\phi_0^p(y_2)\,\Big], \phi_0^p(y_3)\,\Bigg]
\nonumber\\&&\quad =
i \Bigg[\Big\{[ \Delta(x-y_1)\, \phi_0(y_1)^{p-1} ,
\phi_0^p(y_2)\,\Big]\Big\}_{s1}, \phi_0^p(y_3)\,\Bigg]
\nonumber\\&&\quad =
i^2 \Big[\Big\{\Delta(x-y_1)\, \phi_0 ^{p-2} (y_1)\,
\{\Delta(y_1-y_2)\, 
\phi_0^{p-1}(y_2) \}_{s2} \Big\}_{s1}, 
\phi_0^p(y_3)\,\Big]
\nonumber\\&&\quad =
i^2 \Big\{ \Big[ \Delta(x-y_1)\, 
\phi_0^{p-2} (y_1)\Delta(y_1-y_2)
\phi_0^{p-1}(y_2) \}_{s2},\phi_0^p(y_3)\Big]\Big\}_{s1}
\nonumber\\&&\quad =
i^2 \Big\{ \Big[\Delta(x-y_1) \,\phi_0^{p-2} (y_1)\,,
\phi_0^p(y_3)\Big] \,
\{\Delta(y_1-y_2) \, \phi_0^{p-1}(y_2) \}_{s2} \Big\}_{s1}
\nonumber\\&&\qquad \qquad +
i^2 \Big\{ \Delta(x-y_1) \phi_0^{p-2} (y_1) 
\Big[ \{\Delta(y_1-y_2)
\phi_0^{p-1}(y_2) \}_{s2}\,,\phi_0^p(y_3)\,\Big] \Big\}_{s1}
\nonumber\\&&\quad =
2i^3  \Bigg\{\Delta(x-y_1)\, \phi_0^{p-3} (y_1)
\{\Delta(y_1-y_3) \, \phi_0^{p-1}(y_3)\}_{s3}
\{\Delta(y_1-y_2) \, \phi_0^{p-1}(y_2) \}_{s2} \Bigg\}_{s1}
\nonumber\\&&\qquad  +
i^3 \Bigg\{ \Delta(x-y_1)\, \phi_0^{p-2} (y_1) 
\Big\{ \Delta(y_1-y_2)\, \phi_0^{p-2}(y_2) 
 \{ \Delta(y_2-y_3)\, \phi_0^{p-1} (y_3)\}_{s3}
\Big \}_{s2} \Bigg\}_{s1}\,,
\qquad\qquad 
\eeaq
where in the last identity, the factor 2 comes 
from the symmetry of $y_2$ and $y_3$ 
in the symmetrization.
Next, the time-ordered step-function is evaluated as
\beaq
&&\!\!\! \!\!\!
\theta_{123} \, 
\Bigg[\Big[ [\phi_0(x) , \phi_0^p(y_1)], 
\phi_0^p(y_2)\,\Big], \phi_0^p(y_3)\,\Bigg]
\cr 
&&\!\!\!\!\!\!
\quad  = i^3 (\theta_{123} + \theta_{132} )\,
\Bigg\{\Delta(x-y_1)\, \phi_0 ^{p-3} (y_1)\,
\{\Delta(y_1-y_3)\, \phi_0^{p-1}(y_3)\}_{s3}\,
\{\Delta(y_1-y_2)\, \phi_0^{p-1}(y_2) 
\}_{s2}\, \Bigg \}_{s1} 
\cr 
&&\!\!\!\!\!\!
\qquad \qquad +
i^3 \theta_{123} \,
\Bigg\{ \Delta(x-y_1)\, \phi_0^{p-2} (y_1) \,
\Big\{ \Delta(y_1-y_2)\, \phi_0^{p-2}(y_2)\,
\{ \Delta(y_2-y_3) \, \phi_0^{p-1} (y_3)\}_{s3}\,
\Big \}_{s2} \,\Bigg \}_{s1} 
\cr
&&\!\!\!\!\!\!
\quad=
i^3 \Bigg\{\Delta(x-y_1)\, \phi_0^{p-3} (y_1) \,
\{\Delta_{\rm ret} (y_1-y_3)\, \phi_0^{p-1}(y_3)\}_{s3}
\,\{ \Delta_{\rm ret} (y_1-y_2)\, 
\phi_0^{p-1}(y_2) \}_{s2}\,\Bigg\}_{s1}
\cr
&&\!\!\!\!\!\!
\qquad +
i^3 \Bigg\{ \Delta(x-y_1)\, \phi_0^{p-2} (y_1) \,
\Big\{ \Delta_{\rm ret} (y_1-y_2)\,\phi_0^{p-2}(y_2)\,
 \{ \Delta_{\rm ret} (y_2-y_3)\, \phi_0^{p-1} (y_3)\}_{s3}\,
\Big\}_{s2}\,\Bigg \}_{s1}\qquad
\eeaq
where we use the symmetric property of $y_2$ and $y_3$
in the first identity and the step-function identity
in the last identity,
\bea
\theta_{123} + \theta_{132} 
= \theta_{12}\,\theta_{13}\,.
\eea
Again the minimal realization is used
to get the retarded Green's function.
Using the definition of $\phi_1(y)$ and $\phi_2(y)$ in 
(\ref{phiout}), 
and after applying the $\star$-product
we have the out-field of order $g^3$ as
\beaq
\Phi_{(3)} =
\frac g {p!} \int\, dy \,\,
{\cal F}_y
\Bigg(\Big\{ \Delta(x-y)\, 
\phi_0^{p-2}(y) \,\phi_2(y) \Big\}_s
+ \Big \{ \Delta(x-y)\,
\phi_0^{p-3}(y)\, \phi_1^2(y) \Big\}_s
\Bigg)
=\varphi_{(3)}\qquad
\eeaq
which is the out-field given in (\ref{varphiout}).

Higher order proof goes similarly. We provide
up to  the order of $g^4$
since non-local Yukawa theory \cite{non-local}
gives non-trivial
result at this order.
The out-field at the order of $g^4$ is given as
\beaq
&&\!\!\!\!\!\!\!\!\!\!\!\!\!\!\!\!\!\!\!\!\!\!
\Phi_{(4)}(x) =
\Big(\phi_0(x)\, A_4 
- A_1^\dagger\, \phi_0 (x) \, A_3 
+ A_2 ^\dagger\, \phi_0(x)\, A_2 
- A_3 ^\dagger\, \phi_0(x)\, A_1+ A_4^\dagger \, \phi_0(x)
\Big)
\cr 
&&\!\!\!\!\!\!\!\!\!
\quad =
\int\!\!\int\!\!\int\!\!\int
d^Dy_1\, d^Dy_2\, d^Dy_3 \,d^Dy_4 \,\,
{\cal F}_{1234}\,\,
\Bigg(  \theta_{1234} \phi_0(x)\, 
{\cal V}(y_1) {\cal V}(y_2) {\cal V}(y_3) 
{\cal V}(y_4)
\cr 
&&\!\!\!\!\!\!\!\!\!
\qquad
- \theta_{234} {\cal V}(y_1)\phi_0(x) {\cal V}(y_2) 
{\cal V}(y_3) {\cal V}(y_4)
+ \theta_{12}  \theta_{34}\, 
{\cal V}(y_2) {\cal V}(y_1) \phi_0(x)
{\cal V}(y_3) {\cal V}(y_4)
\cr 
&&\!\!\!\!\!\!\!\!\!
\qquad
- \theta_{123} {\cal V}(y_3) {\cal V}(y_2) 
{\cal V}(y_1) \phi_0(x) {\cal V}(y_4)
+ \theta_{1234} {\cal V}(y_4) {\cal V}(y_3) 
{\cal V}(y_2) {\cal V}(y_1) \phi_0(x)
\Bigg)
\cr 
&&\!\!\!\!\!\!\!\!\!
\quad=
\int\!\!\int\!\!\int\!\!\int
d^Dy_1\, d^Dy_2 \, d^Dy_3 \, d^Dy_4  \,\,
\cr 
&&\!\!\!\!\!\!\!\!\!
\qquad\qquad\quad\times
{\cal F}_{1234}\,
\Bigg( \theta_{1234} \,\,
\Bigg[\Big[\big[[\phi_0(x) , {\cal V}(y_1)\,]\,, 
{\cal V}(y_2)\big]\,, {\cal V}(y_3)\Big]\,, 
{\cal V}(y_4)\,\Bigg] \Bigg)\,.
\qquad
\eeaq
Evaluation of the commutator
$[\,[\,[\,[\,\phi_0(x) \,, 
{\cal V}(y_1)\,]\,, {\cal V}(y_2)\,]\,, 
{\cal V}(y_3)\,]\,, {\cal V}(y_4)\,]$
can be done in a few steps:
\bea
&&\!\!\!\!\!\!\!\!\!
\Bigg[\,\Big[\,[\,[\,\phi_0(x)\, , \phi_0^p(y_1)\,]\,, 
\phi_0^p(y_2)\,]\,, \phi_0^p(y_3)\,\Big]\, , 
\phi_0^p(y_4)\,\Bigg]
\cr
&&\!\!\!\!\!\!\!\!\!
= i \Bigg\{\,\Big[\,[\,[ \,\Delta(x-y_1)\, 
\phi_0^{p-1}(y_1)  \,,
\phi_0^p(y_2)\,]\,, \,\phi_0^p(y_3)\,]\,, 
\phi_0^p(y_4)\,\Big]\,\Bigg\}_{s1}
\cr
&&\!\!\!\!\!\!\!\!\!
= i^2 
\Bigg\{\Big\{ \Big[ [\Delta(x-y_1) \phi_0^{p-2}(y_1)  \Delta(y_1-y_2)
\phi_0^{p-1}(y_2),\phi_0 ^p (y_3)\,] ,\phi_0^p(y_4)\Big] 
\Big\}_{s2} \Bigg\}_{s1}
\cr
&&\!\!\!\!\!\!\!\!\!
= i^3
\Bigg\{\,\Big\{\, \Big\{\Big[\,
\Big( 2 \Delta(x-y_1)\, \phi_0 ^{p-3}(y_1)\,
\Delta(y_1 - y_3)\, \phi_0^{p-1}(y_3)\,
\Delta(y_1-y_2) \,\phi_0^{p-1}(y_2) 
\cr
&&\!\!\!\!\!\!\!\!\!
\qquad 
+  \Delta(x-y_1)\, \phi_0^{p-2}(y_1)\,
\Delta(y_1-y_2)\, \phi_0^{p-2}(y_2)\,
\Delta(y_2 - y_3)\, \phi_0^{p-1}(y_3)\,
\Big)\, , \phi_0^p(y_4)\,\Big]\,  
\Big\}_{s3}\,\Big\}_{s2}\,\Bigg\}_{s1}
\cr
&&\!\!\!\!\!\!\!\!\!
=6 \Bigg\{
\Delta(x-y_1) \phi_0^{p-4}(y_1)
\Big \{\Delta(y_1 -y_4)  \phi_0^{p-1}(y_4)\Big\}_{s4}
\Big\{\Delta(y_1 - y_3)\phi_0^{p-1}(y_3)\Big\}_{s3}
\Big\{ \Delta(y_1-y_2) \phi_0^{p-1}(y_2) \Big\}_{s2}
\Bigg \}_{s1}
\cr
&&\!\!\!\!\!\!\!\!\!
\quad+
\Bigg\{\Delta(x-y_1) \phi_0^{p-3}(y_1)
\Big \{\Delta(y_1 - y_3)\phi_0^{p-2}(y_3)
\Big\{\Delta(y_3 -y_4)  \phi_0^{p-1}(y_4) \Big\}_{s4} 
\Big \}_{s3}
\Big\{ \Delta(y_1-y_2) \phi_0^{p-1}(y_2) \Big\}_{s2}
 \Bigg\}_{s1}
\cr
&&\!\!\!\!\!\!\!\!\!
\quad
+ \Bigg\{ \Delta(x-y_1) \phi_0^{p-3}(y_1)
\Big\{ \Delta(y_1 - y_3)\phi_0^{p-1}(y_3)\Big\}_{s3}
\Big\{ \Delta(y_1-y_2) \phi_0^{p-2}(y_2)
\{ \Delta(y_2 -y_4)  \phi_0^{p-1}(y_4)
\}_{s4} \Big\}_{s2}\Bigg \}_{s1}
\cr
&&\!\!\! \!\!\!\!\!\!
\quad+ \Bigg\{ \Delta(x-y_1) \phi_0^{p-3}(y_1)
\{ \Delta(y_1 -y_4)  \phi_0^{p-1}(y_4) \}_{s4}
\Big\{ \Delta(y_1-y_2) \phi_0^{p-2}(y_2)
\Big\{ \Delta(y_2 - y_3)\phi_0^{p-1}(y_3) \Big\}_{s3}
\Big\}_{s2} \Bigg\}_{s1}
\cr
&&\!\!\!\!\!\!\!\!\!
\quad
+  2 \Bigg\{\Delta(x-y_1) \phi_0^{p-2}(y_1)
\Big\{ \Delta(y_1-y_2) \phi_0^{p-3}(y_2)
\{ \Delta(y_2 -y_4)  \phi_0^{p-1}(y_4)
\}_{s4} \Big\}_{s2}
\Big\{ \Delta(y_2 - y_3)\phi_0^{p-1}(y_3)
\Big\}_{s3} \Bigg\}_{s1}
\cr
&&\!\!\!\!\!\!\!\!\!
\quad
+ \Bigg \{ \Delta(x-y_1) \phi_0^{p-2}(y_1)
\Big\{ \Delta(y_1-y_2) \phi_0^{p-2}(y_2)
\{ \Delta(y_2 - y_3)\phi_0^{p-2}(y_3)
\}_{s3} \Big\}_{s2}
\Big\{\Delta(y_3 -y_4)  \phi_0^{p-1}(y_4)
\Big\}_{s4} \Bigg\}_{s1}\,.
\eea
\eject
Using the identities
\bea
&& \theta_{1234} + (234\, \rm{permutation})
= \theta_{12} (\theta_{234}+ \theta_{243})
+ \theta_{13} (\theta_{324}+\theta_{342})
+ \theta_{14}(\theta_{423}+\theta_{432})
\\&&\qquad\qquad\qquad
=  \theta_{12}\theta_{23}\theta_{24}
+ \theta_{13}\theta_{32}\theta_{34}
+\theta_{14}\theta_{42}\theta_{43}
=\theta_{12}\theta_{13} \theta_{14}
\\ &&
\theta_{1234} + \theta_{1324} + \theta_{1342}
= (\theta_{12}\theta_{13} - \theta_{13}\theta_{32}) \theta_{34}
+ \theta_{1324} + \theta_{1342}
\\&& \qquad\qquad\qquad
 = \theta_{12}\theta_{13} \theta_{34}
- \theta_{132} \theta_{34}
+ \theta_{1324} + \theta_{1342}
= \theta_{12}\theta_{13} \theta_{34}
\\&&
\theta_{1234}+ \theta_{1243} = \theta_{12}\theta_{23}\theta_{24} \,,
\eea
we may put the commutator with the time-ordering as
\bea
&&\!\!\!\!\!\!\!\!\!
\theta_{1234} \,
[[[[\phi_0(x) , \phi_0^p(y_1)], \phi_0^p(y_2)]
, \phi_0^p(y_3)], \phi_0^p(y_4)]
= - \int d^Dy\, \Delta(x-y) \Bigg\{
\delta^D(y-y_1) \,\phi_0^{p-4}(y_1)
\cr
&&\!\!\!\!\!\!\!\!\!
\times
\Bigg(
\{\Delta_{\rm ret}(y_1 -y_4)  \phi_0^{p-1}(y_4)\}_{s4}\,
\{\Delta_{\rm ret}(y_1 - y_3)\phi_0^{p-1}(y_3)\}_{s3}\,
\{ \Delta_{\rm ret}(y_1-y_2) \phi_0^{p-1}(y_2)\}_{s2}
\cr
&&\!\!\!\!\!\!\!\!\!
\quad+
\phi_0(y_1)
\Big \{\Delta_{\rm ret}(y_1 - y_3)\phi_0^{p-2}(y_3)
\{\Delta_{\rm ret} (y_3 -y_4)  \phi_0^{p-1}(y_4) \}_{s4} \Big \}_{s3}
\{ \Delta_{\rm ret} (y_1-y_2) \phi_0^{p-1}(y_2) \}_{s2}
\cr
&&\!\!\!\!\!\!\!\!\!
\quad
+  \phi_0^2(y_1)
\Big\{ \Delta_{\rm ret} (y_1-y_2) \phi_0^{p-3}(y_2)
\{ \Delta_{\rm ret} (y_2 -y_4)  \phi_0^{p-1}(y_4)\}_{s4} \Big\}_{s2}
\{ \Delta_{\rm ret} (y_2 - y_3)\phi_0^{p-1}(y_3)\}_{s3}
\cr
&&\!\!\!\!\!\!\!\!\!
\quad
+ \phi_0^2(y_1)
\Big\{ \Delta_{\rm ret} (y_1-y_2) \phi_0^{p-2}(y_2)
\{ \Delta_{\rm ret} (y_2 - y_3)\phi_0^{p-2}(y_3)\}_{s3} \Big\}_{s2}
\{\Delta_{\rm ret} (y_3 -y_4)  \phi_0^{p-1}(y_4) \}_{s4}
\Bigg)  \Bigg\}_{s1}\,.
\eea

Using the definitions of $\phi_n (y)$ and 
after applying the $\star$-product
we have the out-field of order $g^4$ as
\beaq
\Phi_{(4)} &=& \frac g {p!} 
\int d^Dy \,{\cal F}_y \Bigg\{\,
\Delta(x-y)\,\,
\Big( \phi_0^{p-4}(y) \phi_1^3(y)
+ \phi_0^{p-3} (y)\phi_1 (y)\phi_2 (y)
+ \phi_0^{p-2} (y)\phi_3 (y) \Big)\Bigg\}_{s(y)}
\nonumber\\&=&
\varphi_{(4)}\,.
\eeaq
As demonstrated in the above derivation, the minimal realization
and the $\star$-time ordering
are enough for proving that the S-matrix connects the
in- and out-field correctly to all orders of perturbation.

\vskip 0.5cm
\subsection{Unitarity of S-matrix}
\noindent

In this section, we provide a proof
that this S-matrix is unitary,
\beq
S S^\dagger = S^\dagger S =1\,.
\eeq
To do this we use the S-matrix
in Eq.~(\ref{S-matrix}) and (\ref{An}),
the coupling constant expanded version of $S$-matrix,
and evaluate  $S S^\dagger $ order by order in $g$.
We remark that the product of $S S^\dagger $
is not the $\star$-product but is the ordinary product
since $S$-matrix does not depend on coordinates explicitly.

The unitarity at the order of $g$ is
trivially satisfied since $ A_1 ^\dagger = A_1$.
At the order of $g^2$,
the unitarity condition is given as
\beq
A_2 + A_2^\dagger = A_1^\dagger A_1\,.
\eeq
The proof goes as follows:
\beaq
{\rm LHS} &=& A_2 + A_2^\dagger
\nonumber \\ &=&
\int\!\int
d^Dy_1\, d^Dy_2 {\cal F}_{12} 
\Bigg( \theta_{12}
\Big( {\cal V}(\phi_0(y_1))\,
{\cal V}(\phi_0(y_2))
+ {\cal V}(\phi_0(y_2))\,
{\cal  V}(\phi_0(y_1))\Big) \Bigg)
\nonumber\\
&=&
\int\!\int
d^Dy_1 \,d^Dy_2 \, {\cal F}_{12}
\Big( (\theta_{12} +\theta_{21}) 
{\cal V}(\phi_0(y_1))\, 
{\cal V}(\phi_0(y_2)) \Big)
\nonumber\\
&=&
\int d^D y_1 {\cal F}_{1}
\Big( {\cal V} (\phi_0(y_1)) \Big)
\int d^D y_2 {\cal F}_{1}
\Big( {\cal V}(\phi_0(y_2)) \Big)\,,
\nonumber\\
{\rm RHS} &=& A_1^\dagger A_1
\nonumber\\
&=&
\int d^D y_1 {\cal F}_{1}
\Big( {\cal V}(\phi_0(t_1)) \Big)
\int d^Dy_2 {\cal F}_{1}
\Big( {\cal V}(\phi_0(t_2)) \Big)\,
\eeaq
and therefore, $LHS= RHS$.
(Note that the $\dagger$ operation is applied
to the fields $\phi_0 $'s not the time-ordering
or $\star$-operation).
Here
we use the change of variables to get the third line
and the identity
$\theta_{12}+ \theta_{21} =1$. 
It should be noted that
this step-function identity
always holds even when the coordinates are split
as far as the split coordinates are concerned:
\bea
\theta( x_i-y_j ) + \theta( y_j-x_i ) =1 \,.
\eea
This is the reason
why the unitarity holds without using the minimal realization.

At the order of $g^3$,
the unitarity condition is given as
\beq
A_3 - A_3^\dagger = A_1^\dagger A_2 -  A_2^\dagger A_1  \,.
\eeq
The proof goes as follows:
\beaq
{\rm LHS} &=& A_3 - A_3^\dagger
\nonumber\\ &=&
\int\!\!\int\!\!\int
d^Dy_1\, d^Dy_2\, d^Dy_3 \,\,
{\cal F}_{123} \Bigg( \theta_{123}
\Big( {\cal V}(\phi_0(y_1))\, 
{\cal V}(\phi_0(y_2))\, {\cal V}(\phi_0(y_3))
\nonumber\\&&
\qquad\qquad\qquad\qquad\qquad\qquad
- {\cal V}(\phi_0(y_3))\, 
{\cal V}(\phi_0(y_2))\, 
{\cal V}(\phi_0(y_1))\Big) \Bigg)
\nonumber\\ &=&
\int\!\!\int\!\!\int
d^Dy_1\, d^Dy_2\, d^Dy_3 \,\,
{\cal F}_{123} \Bigg(
\Big(\theta_{123} -\theta_{321} \Big)
{\cal V}(\phi_0(y_1)) 
{\cal V}(\phi_0(y_2)) 
{\cal V}(\phi_0(y_3)) \Bigg)
\nonumber\\ &=&
\int\!\!\int\!\!\int
d^Dy_1\, d^Dy_2 \, d^Dy_3\,\,
{\cal F}_{123}
\Bigg( \Big( \theta_{23} -\theta_{21} \Big)
{\cal V}(\phi_0(y_1))\,
{\cal V}(\phi_0(y_2))\,
{\cal V}(\phi_0(y_3))\Bigg)
\nonumber\\ &=&
\int\!\!\int\!\!\int
d^Dy_1\, d^Dy_2\, d^Dy_3\,\,
\Bigg( {\cal F}_{1} \Big( {\cal V}(y_1)\Big ) \,
{\cal F}_{23} \Big( \theta_{23} 
{\cal V}(y_2) \, {\cal V}(y_3)\Big)
\nonumber\\&&
\qquad\qquad\qquad\qquad\qquad\qquad
-\, {\cal F}_{12} \Big( \theta_{21} 
{\cal V}(y_1) {\cal V}(y_2) \Big)
\,{\cal F}_{3} \Big( {\cal V}(y_3) \Big) \Bigg)
\eeaq
\beaq
{\rm RHS} &=& A_1^\dagger A_2 -  A_2^\dagger A_1
\nonumber \\ &=&
\int\!\!\int\!\!\int
d^Dy_1\, d^Dy_2 \, d^Dy_3  \Bigg(
{\cal F}_{1} \Big( {\cal V}(y_1) \Big)
{\cal F}_{23} \Big( \theta_{23}{\cal V}(y_2)\, {\cal V}(y_3)
\Big)
\nonumber\\&&
\qquad\qquad\qquad\qquad\qquad\qquad
- {\cal F}_{12} \Big(\theta_{12}\,
{\cal V}(y_2)\, {\cal V}(y_1) \Big)
{\cal F}_{3} \Big({\cal V}(y_3) \Big)
\Bigg) \,,
\eeaq
where we use the identity
\beq
\theta_{123}- \theta_{321}
= (1-\theta_{21}) \theta_{23} -\theta_{32} \theta_{21}
= \theta_{23}-\theta_{21} \,.
\eeq Comparing with both sides, we have
${\rm LHS}= {\rm RHS}\,$.

At the order of $g^4$,
the unitarity condition is given as
\beq
A_4+ A_4^\dagger
= A_1^\dagger A_3 -  A_2^\dagger A_2 +A_3^\dagger A_1  \,.
\eeq
The proof goes as follows:
\beaq
&&\!\!\!\!\!\!\!\!\!
{\rm LHS} = A_4 + A_4^\dagger
\cr
&&\!\!\!\!\!\!\!\!\!
=
\int d^Dy_1\, \cdots\, \!\!\int d^Dy_4\,   
{\cal F}_{1234} \Bigg( \theta_{1234}\,
\Big({\cal V}(y_1)\, {\cal V}(y_2)\,
{\cal V}(y_3)\, {\cal V}(y_4)
\cr 
&& \qquad\qquad\qquad\qquad\qquad\qquad\qquad
+ {\cal V}(y_4)\, {\cal V}(y_3)\,
{\cal V}(y_2)\, {\cal V}(y_1) \,\Big) \Bigg)
\cr
&&\!\!\!\!\!\!\!\!\!
=
\int d^Dy_1\, \cdots\, \!\!\int d^Dy_4\,   
{\cal F}_{1234} \Bigg(
\Big(\theta_{1234} + \theta_{4321} \Big)\,
{\cal V}(y_1)\, {\cal V}(y_2)\,
{\cal  V}(y_3)\, {\cal V}(y_4) \Bigg)\,.\quad
\eeaq
Using the identities, 
\bea 
&& \theta_{1234} = \theta_{234}
-( \theta_{2134} + \theta_{2314} + \theta_{2341} )
=\theta_{234}-(\theta_{2134} + \theta_{23}\theta_{31} \theta_{34})
\\ && \theta_{4321}=\theta_{321}
-(\theta_{3421} + \theta_{3241}+\theta_{3214})
=\theta_{321}-( \theta_{3421} +  \theta_{32} \theta_{21} \theta_{24})
\\ &&
\theta_{2134} +\theta_{23}\theta_{31} \theta_{34}
=  \theta_{34} (\theta_{213}+ \theta_{231})
= \theta_{34} \theta_{21} \theta_{23}
\\ &&  \theta_{3421} +\theta_{32} \theta_{21} \theta_{24}
= \theta_{21} ( \theta_{342} + \theta_{324})
= \theta_{21} \theta_{34} \theta_{32} \,,
\eea
we have
\beq
{\rm LHS} =
\int d^Dy_1\, \cdots\, \!\!\int d^Dy_4\,   
{\cal F}_{1234} \Bigg(
\Big( \theta_{234}
-\theta_{21}  \theta_{34}
+ \theta_{321} \Big)\,
{\cal V}(y_1)\, {\cal V}(y_2)\,
{\cal V}(y_3)\, {\cal V}(y_4)\,\Bigg)\,.
\eeq
On the other hand,
\beaq
&&\!\!\!\!\!\!\!\!\!
{\rm RHS}
= A_1^\dagger A_3 -  A_2^\dagger A_2 +A_3^\dagger A_1
\cr
&&\!\!\!\!\!\!\!\!\!
= \int d^Dy_1\, \cdots\, \!\!\int d^Dy_4\,   
\Bigg( {\cal F}_{1} \Big( {\cal V}(y_1) \Big)
{\cal F}_{234} \Big( \theta_{234} \,
{\cal V}(y_2)\, {\cal V}(y_3)\,
{\cal V}(y_4)\, \Big)
\cr 
&& \qquad\qquad\qquad\qquad \qquad
- {\cal F}_{12} \Big(\theta_{12}\, 
{\cal V}(y_2)\, {\cal V}(y_1) \Big)
{\cal F}_{34} \Big( \theta_{34} \,
{\cal V}(y_3)\, {\cal V}(y_4))\, \Big)
\cr 
&& \qquad\qquad\qquad\qquad \qquad
+ {\cal F}_{123} \Big( \theta_{123} \,
{\cal V}(y_3)\, {\cal V}(y_2)\,
{\cal V}(y_1)\,  \Big)
{\cal F}_{4} \Big( {\cal V}(y_4)\, \Big)
\Bigg)
\cr
&&\!\!\!\!\!\!\!\!\!
=
\int d^Dy_1\, \cdots\, \!\!\int d^Dy_4\,   
\Bigg( {\cal F}_{1234} \Bigg(
\Big( \theta_{234}
-\theta_{21}  \theta_{34}
+ \theta_{321} \Big)\,
{\cal V}(y_1)\, {\cal V}(y_2)\,
{\cal V}(y_3)\, {\cal V}(y_4)\, \Bigg) \Bigg)\,.
\qquad
\eeaq
Comparing with both sides, we have LHS = RHS.

One can confirm that the higher order proof goes similarly with
the ordinary perturbation case.
In this proof,  only the time-ordering matters
irrespective of the $\star$-operation.

One may put the time-ordering out-side of the
star-operation as far as the unitarity is concerned.
However, the time-ordering outside the star-operation
does not fulfill the correct in- and out-field relation.
It is remarked that
the time-ordering outside the
$\star$-operation is different
from the time-ordering inside the
$\star$-operation up to higher derivatives.
It is like the contact terms in the ordinary gauge theory.

To see this we give an explicit expression for this difference
up to order of $g^3$.
At the order of $g$, there is no distinction
between two since there is no
time ordering.  At the order of $g^2$,
let us denote the ordinary time-ordered one as $a_2$,
which puts the time-ordering outside the
$\star$-operation:
\beq
a_2 = \int\!\!\int dy_1 dy_2 \,
\theta_{12} {\cal F}_{12} \Big( V (t_1) V(t_2) \Big)\,.
\eeq
$a_2$ satisfies the relation:
\bea
a_2 + a_2^\dagger  =A_1^2\,.
\eea
The difference is denoted as $c_2$:
\beq
ic_2 =A_2 -a_2\,,\qquad c_2 = c_2^\dagger \,,
\eeq
which is given as
\beaq
i c_2 &=&
-\frac 12
\int\!\!\int  d^Dy_1 \, d^Dy_2\,
\Bigg(
\theta_{12}\, {\cal F}_{12}
-{\cal F}_{12} \,\theta_{12}\Big)
\Big( {\cal V}(y_1)\, {\cal V}(y_2) + {\cal V}(y_1)\, {\cal V}(y_2)\Big)
\nonumber\\&&\quad
-\frac 12
\int\!\!\int  d^Dy_1\, d^Dy_2\,
\Bigg(
\theta_{12}\, {\cal F}_{12}
-{\cal F}_{12} \,\theta_{12}\Big)
\Big( [\, {\cal V}(y_1), {\cal V}(y_2)\,]
\Big)
\nonumber\\
&=&
-\frac 12
\int\!\!\int d^Dy_1\, d^Dy_2\,
\Bigg(
(\theta_{12} + \theta_{21})\, {\cal F}_{12}
-{\cal F}_{12} \,(\theta_{12} + \theta_{21}) \Big)
\Big( {\cal V}(y_1)\, {\cal V}(y_2) \Big)
\nonumber\\&&\qquad
-\frac 12
\int\!\!\int d^Dy_1\, d^Dy_2\,
\Bigg(  \theta_{12}\, {\cal F}_{12}
-{\cal F}_{12} \,\theta_{12}\Big)
\Big( [\, {\cal V}(y_1)\,, {\cal V}(y_2)\,]  \Big)
\nonumber\\
&=&-\frac12  \int\!\!\int 
dy_1 dy_2  \, \Big(\theta_{12}\, {\cal F}_{12}
- {\cal F}_{12}  \, \theta_{12}\Big) \,
\Big( \, [\, {\cal V}(y_1)\,, {\cal V}(y_2)\, ]\Big) \,,
\eeaq
where we use the identity 
$\theta_{12} + \theta_{21}=1$.
This is the source of higher derivative terms
to the lowest order,
which is to be
supplemented by the S-matrix proposed in \cite{bahns}.
If one evaluates the commutator of the step function
and the $\star$-product,
this leaves us with the time derivatives
of the fields and of the spectral functions.

For the order of $g^3$, we have
\beq
A_3 = a_3 + i c_2 A_1 + c_3\,.
\eeq
$a_3$ is the ordinary time-ordered one:
\beq
a_3 =\int\!\!\int\!\!\int
d^Dy_1\, d^Dy_2\, d^Dy_3\,
\theta_{123} {\cal F}_{123} 
\Big( {\cal V}(y_1)\, {\cal V}(y_2)\,
{\cal V}(y_3)\Big)\,.
\eeq
Using the identity,
\bea
&&\int\!\!\int\!\!\int
d^Dy_1\, d^Dy_2\, d^Dy_3\,
\theta_{123}  {\cal V}(y_1)\, {\cal V}(y_2)\,
{\cal V}(y_3)
\\&&\qquad = \frac 16
\int\!\!\int\!\!\int
d^Dy_1\, d^Dy_2\, d^Dy_3\,
{\cal V}(y_1)\, {\cal V}(y_2)\, {\cal V}(y_3)
\\ &&\qquad \qquad
+ \frac13
\int\!\!\int\!\!\int
d^Dy_1\, d^Dy_2\, d^Dy_3\,
(\theta_{123} + \theta_{132}) \,
[\,{\cal V}(y_1),[\,{\cal V}(y_2), {\cal V}(y_3)\,]\,]
\\&&\qquad \qquad
+ \frac12
\int\!\!\int\!\!\int
d^Dy_1\, d^Dy_2\, d^Dy_3\,
\theta_{12}
[\,[\,{\cal V}(y_1)\,,{\cal V}(y_2)\,]\,,
{\cal V}(y_3)\,]
\eea
we have
\beq
c_3 =-\frac13
\int\int\int
d^Dy_1\, d^Dy_2\, d^Dy_3\,
\Big(\theta_{123}\, {\cal F}_{123}
-{\cal F}_{123} \,\theta_{123}\Big) \,
\Big( \,[{\cal V}_1, [{\cal V}_2, {\cal V}_3]]
+ [{\cal V}_2, [{\cal V}_1 , {\cal V}_3]] \Big)
\,,
\eeq
with $ c_3 = c_3^\dagger$.

\section{Feynman rule in Momentum-space \label{f}}
\noindent

In this section we illustrate the perturbation approach to the STNC field theory
in the momentum-space.
The momentum space calculation will be complementary
to the coordinate space representation
described in section \ref{s}.
The minimal realization of the time-ordering is 
to be properly represented. 
For definiteness, we consider  $\phi^4$ theory,
\beq  L_I (t)
= -\frac{\lambda}{4!} \,\int d^{D-1}x \,
\phi_\star^4 (x) \,.
\eeq

Two-point function is represented 
in terms of the positive spectral function
$\Delta_+ (x)$  instead of Feynman propagator,
\bea
\Delta_+ (x)
= \langle 0\mid \phi_{\rm in} (x)
\phi_{\rm in} (0) \mid 0 \rangle
=\int \frac{d^D k}{(2\pi)^D} \,\, e^{-ikx}
\,\,\tilde \bigtriangleup_+ (k)
\eea
where $ \tilde \Delta_+ (k)$
is the Fourier transform of the free spectral function,
\beq
\tilde \bigtriangleup_+ (k)
= \begin{picture}(70,23)(-10,10)
\put(05,10){\line(1,0){55}}
\put(25,12){\footnotesize{$k$}}
\put(30,8){\footnotesize{$>$}}
\put(5,10){\circle{2}}
\put(60,10){\circle{2}}
\end{picture}
= 2\pi \delta (k^2 - m^2 ) \theta ( k^0)
\label{k-2pt}
\eeq
where we specify the arrow to denote the momentum flow.

In addition, we need a ``time-ordered''
spectral function $\Delta_R(x)$
to describe the time ordering effect.
\beaq
\Delta_R (x) &=&  \theta(x^0) \Delta_+ (x)
= \int \frac{d^D k}{(2\pi)^D} \,\, e^{-ikx}
\,\,\tilde \Delta_R (k)
\nonumber \\
\tilde \Delta_R (k) &=&
\frac{i}{2\omega_k} \frac1{(k_0- \omega_k + i\epsilon)}
= \begin{picture}(70,23)(-20,10)
\put(05,10){\line(1,0){55}}
\put(30,7){ $\triangleright$}
\put(20,12){\footnotesize{$k$}}
\put(5,10){\circle{2}}
\put(60,10){\circle{2}}
\end{picture}
\label{k-dir-2pt}
\eeaq
with $\omega_k = \sqrt{\vec k^2 + m^2}$.
$\tilde \Delta_R (k)$ is represented as a triangled arrow
to emphasize the ordering effect.

It is noted that the time-ordered two-point function
$\Delta_R (x)$ in (\ref{k-dir-2pt})
is not confused with the retarded Green's function 
$\Delta_{\rm ret} (x)$ 
given in (\ref{ad-ret}): Each has a different pole structure.
The Feynman propagator is given in terms of the
time-ordered spectral function,
\bea
i \Delta_F (x) = \Delta_R (x)  + \Delta_R (-x) \,.
\eea

The four-point vertex is given as
\bea
-i (2\pi)^d \delta^d (p_1 + p_2 +p_3 +p_4)
\,\Gamma_4 (k_1, k_2, k_3, k_4)
\eea
and its lowest order diargam is given as
\bea
-i \Gamma_4^{(0)} (k_1, k_2, k_3, k_4)
&=& \begin{picture}(60,30) (0,30)
\put(-8,50){\footnotesize{$p_3$}}
\put(50,50){\footnotesize{$p_4$}}
\put(0,20){\line(2,1){50}}
\put(0,45){\line(2,-1){50}}
\put(36,38){\line(1,0){5}}
\put(36,38){\line(0,1){4}}
\put(8,38){\line(1,0){5}}
\put(13,38){\line(0,1){4}}
\put(8,26.8){\line(1,0){5}}
\put(13,22.5){\line(0,1){3}}
\put(37,26.5){\line(1,0){5}}
\put(37,23){\line(0,1){3}}
\put(25,32.5){\circle*{2}}
\put(-8,10){\footnotesize{$p_1$}}
\put(50,10){\footnotesize{$p_2$}}
\end{picture}
=   -i \lambda\, v(p_1,p_2,p_3,p_4)
\eea
where
\bea
 v(p_1,p_2,p_3,p_4)
&=&
\frac 13 \Bigg(
\cos\Big( \frac {p_1 \wedge p_2}2 \Big)
\cos\Big(\frac {p_3 \wedge p_4}2 \Big)
\\ && \quad
+
\cos\Big(\frac {p_1 \wedge p_3}2 \Big)
\cos\Big(\frac {p_2 \wedge p_4}2 \Big)
+
\cos\Big(\frac {p_1 \wedge p_4}2 \Big)
\cos\Big(\frac {p_2 \wedge p_3}2 \Big)
\Big)\,.
\eea
The vertex function $v$ is permutationally symmetric in the
external momentum indices
and is insensitive to the sign of the momenta;
\bea
v(p_1,p_2,p_3,p_4)
=v(\pm p_1, \pm p_2, \pm p_3, \pm p_4)
= v(p_{\sigma(1)} ,p_{\sigma(2)}, p_{\sigma(3)}, p_{\sigma(4)} )
\eea
where $\sigma(i)$ is the permutation operation.

\vskip 0.2cm
The Feynman rule for this theory is summarized as follows. \\
(1) Each vertex is assigned as 
$-i\lambda v(p_1,p_2,p_3,p_4)$ where $p_i$'s
are incoming momenta of four legs 
and its total momentum vanishes.\\
(2) The legs are either external legs 
or are connected to other
vertices, making internal lines. \\
(3) Each vertex is numbered so
that the internal lines are assigned with arrows. 
The arrows point
from high-numbered vertex to low-numbered one. \\
(4) Among the arrows, 
only one arrow between two adjacent vertices is assigned
as triangled one and the total number 
of the triangled arrows should be $n-1$ 
for the $n $ connected vertices. This assignment
is due to the minimal realization of the time-ordered
step function. \\
(5) The diagrams with the same
topology with arrows are identified 
and the numbering of vertices
is ignored. As a result the number of diagrams 
are reduced from the original $n!$ diagrams. \\
(6) The distinctive Feynman diagrams
are multiplied with the symmetric factors. \\
(7) The momentum flows along the arrows. 
The arrowed internal line with momentum $k$ is
assigned as $\tilde \Delta_+ (k)$ 
and the triangled arrowed
internal line as $\tilde \Delta_R (k)$.

\vskip 0.2cm
Note that the rule (5) originates from the time-ordering. To
understand this, we provide a few examples. Let us denote the
three vertex diagram a---b---c as the numbered vertex 
$(a,b,c)$, where as many as internal lines between vertices
may exist.  
When two diagrams numbered as $(3,1,2)$ and $(2,1,3)$ are 
combined,
\bea \theta_{123} \Big( (3,1,2) + (2,1,3)\Big) = \Big(
\theta_{123} +\theta_{132} \Big) (3,1,2) = \Big( \theta_{123}
+\theta_{132} \Big) (2,1,3)
 \,, \eea
one may use the step function identity 
$ \theta_{123} +\theta_{132} = \theta_{12}\, \theta_{13}
\,,\,\,$ 
and rearrange the time-ordering 
so that the ordering is directly relevant 
for the diagram $(213)$ or $(312)$: 
\bea 
\theta_{123} \Big(
(3,1,2) + (2,1,3) \Big) = \theta_{12}\, 
\theta_{13} \, (2,1,3) = \theta_{12}\, \theta_{13} \, (3,1,2) \,. 
\eea 
This reduces the two numbered diagrams 
into the one distinct Feynman diagram with the
arrow topology; \/  \hbox{o---$\!\!\!
>\!\!$---o---$\!\!\! <\!\!$---o}\/ .
\vskip 0.1cm

Consider a four-point vertex diagram denoted as $\{abcd\}\equiv$
\hbox{ \rlap {\raise0pt
    \hbox{ b ----- c ----- d\/ . }}
\rlap{ \lower10pt\hbox{$\hskip1.cm |$ }} \rlap{ \lower22pt\hbox{
$\hskip0.58cm $ a  }} } $\qquad\qquad \qquad $ \\ \vskip0.1cm
\noindent
 6 diagrams numbered as  $\{a b 1 c\}$ with $a,
b$, and $c$ the permutations of $234$ will be reduced to a Feynman
diagram \hbox{ \rlap {\raise0pt
    \hbox{ o-----$\!\!>\!\!\!\!$-----$\!\! $
                 o$\!$-----$\!\!\!\!< \!\!\!\!$-----$\!\!
                  $ o , }} \rlap{ \lower10pt\hbox{$\hskip1.24cm |$ }} \rlap{
\lower15pt\hbox{$\hskip0.94cm$ {\footnotesize$\wedge$}  }} \rlap{
\lower22pt\hbox{ $\hskip0.83cm |$   }} } $\qquad \qquad
\qquad\quad$
 since $ \Big(
\theta_{1234} + {\rm permutations} \; {\rm of}\;  234\Big)
=\theta_{12} \theta_{13}\theta_{14} $.

\vskip 0.3cm
Three diagrams $(2134)$, $(3124)$ and $(4123)$ with
$(abcd)\equiv$\hbox{ a-----b-----c-----d} are reduced to a Feynman
diagram
\hbox{ o---$\!\!>\!\!\!$---o---$\!\!\! < \!\!\!$---o%
---$\!\!\! < \!\!\!$---o}\/\/
since $ \theta_{1234} + \theta_{1324} + \theta_{1342} =
\theta_{12}\theta_{13} \theta_{34} \,\,\,$. The reduction of the
numbered diagrams to an arrowed one is very general in momentum
space and hence, the rule (5) follows.

\subsection{Self-energy:}

Self energy is defined as \beq -i \Sigma_{(1)}(p_1, p_2) \,
(2\pi)^D\,  \delta^D (p_1 + p_2) \equiv \langle -\!p_2| \,S-1\,  |
p_1 \rangle_c \eeq where $ \langle \,\cdots\,\rangle_c  $ refers
to the amputated one-particle irreducible function. In
perturbation, we use the one particle state representation with momentum
$p$, $\langle p \,| \,\phi_{\rm in}(x) |0 \rangle =N e^{ipx}$ with
$N=1$ as a proper normalization constant.

The one loop contribution to the self-energy comes from the first
term of S-matrix, $A_1$ in (\ref{S-matrix}); $\langle -\! p_2|\, i
A_1  | p_1 \rangle_c $ \bea -i \Sigma_{(1)}(p_1, p_2) &=&
\frac12
\;
\begin{picture}(70,23)(-10,10)
\put(05,10){\line(3,0){55}}
\put(30,22){\circle{25}}
\put(25.5,31.5){\footnotesize{$>$}}
\put(28,22){\footnotesize{$k$}}
\put(-8,12){\footnotesize{$p_1$}}
\put(62,12){\footnotesize{$p_2$}}
\put(8,7.5){\footnotesize{$>$}}
\put(44,7.5){\footnotesize{$<$}}
\end{picture}
\\ &=&
-\frac {i\lambda} 2 \, \int_k \, 
\tilde\Delta_+(k) \, v(p_1,p_1,k, k) 
\eea 
where $\frac12$ is the symmetric factor and $\int_k$ is
the abbreviated notation for the momentum integration; 
\bea 
\int_k \equiv \int  \frac{d^{D}k}{(2\pi)^{D}} \,. 
\eea 
This one-loop contribution can be written as 
\beq 
\Sigma_{(1)}(p_1, p_2) 
= \frac {\lambda} 2 \, 
\Bigg( \Big(\frac 23 \Big) \,\int_k \,  
\frac i{k^2 -m^2 +i \epsilon} + \int_k \,  
\tilde\Delta_+(k) \, 
\frac{\cos(p_1 \wedge k)}3 \Bigg)\,. 
\label{sigma1} 
\eeq 
Here we use the identity (see Appendix), 
\beq 
\label{appen1}
\int_k \tilde \Delta_+ (k) = \int_k \frac i {k^2
-m^2 +i\epsilon}\,. 
\eeq 
The first term in (\ref{sigma1}) is
UV-divergent when $D\ge 2$ 
and can be absorbed into the mass renormalization. 
Note that the factor 2/3 is different for the the
commuting case. 
 
The second term is the non-planar contribution.
One may put the integration for even~$D$ 
(see Appendix) as 
\beaq 
\label{appen2} 
\int_k \tilde \Delta (k) 
\cos(p \wedge k) 
&=& \int_0^\infty d\alpha \,\frac 1{(4\pi
\alpha)^{D/2}}\, 
e^{-\alpha m^2 - \frac{p\,\circ p}{4 \alpha}}
\nonumber\\
&=& \frac{m^{(D-2)/2}}{(2\pi)^{D-2}}\,
(p\,\circ p)^{\frac{2-D}{4}}\,
K_{\frac{D-2}{2}}\left(m\sqrt{p\,\circ p}\right), 
\eeaq 
where 
$p\circ k =p^\mu \theta_{\mu\nu} \theta
^{\nu\rho}k_\rho\, $ 
and the $K_\nu (x)$ is the modified Bessel function. 
Therefore, the second term is finite as far
as $\theta$ and mass do not vanish. 
The feature that non-planar diagram is finite 
is very general in SSNC QFT \cite{perturb}. 
The same conclusion applies to STNC QFT also.  
Furthermore, it should be noted that
unlike in SSNC QFT, there is no UV-IR mixing since
$p\,\circ\,p \ge  m^2 $ 
when $p$ is on-shell \cite{uv-ir}. 
This feature is not changed even if 
$\Sigma_{(1)}$ is included in a  
higher loop graph since $\Sigma_{(1)}$ 
is connected through the 
two-point function given in (\ref{k-2pt}) 
which maintains the on-shell condition 
due to the delta-function,
and therefore, the loop-diagrams 
do not present any UV-IR mixing problem.

The two-loop contribution comes from the terms:
$\langle -\! p_2|  i^2 A_2  | p_1\rangle_c\,\, $.
\bea
-i \Sigma_{(2)}(p_1, p_2)
= \Sigma_{(2a)}(p_1, p_2) +\Sigma_{(2b)}(p_1, p_2)\,.
\eea
The first contribution is given as
\bea
-i \Sigma_{(2a)}(p_1, p_2)
&=& \frac14\;
\begin{picture}(90,23)(-10,10)
\put(10,10){\line(3,0){55}}
\put(35,20){\circle{20}}
\put(35,40){\circle{20}}
\put(21.5,17){\footnotesize{$\triangle$}}
\put(41.5,17){\footnotesize{$\wedge$} }
\put(31,47.5){\footnotesize{$>$} }
\put(-3,12){\footnotesize{$p_1$}}
\put(67,12){\footnotesize{$p_2$}}
\end{picture}
+  \frac14\;
\begin{picture}(90,23)(-10,10)
\put(10,10){\line(3,0){55}}
\put(35,20){\circle{20}}
\put(35,40){\circle{20}}
\put(21,18){\footnotesize{$\bigtriangledown$}}
\put(41.5,18.5){\footnotesize{$\vee$} }
\put(31,47.5){\footnotesize{$>$} }
\put(-3,12){\footnotesize{$p_1$}}
\put(67,12){\footnotesize{$p_2$}}
\end{picture}
\\&=&
-\frac{\lambda^2}2 \,\int_{k\,,\ell} \tilde{\Delta}_+(\ell)
\,\tilde{\Delta}_R(k)\,\tilde{\Delta}_+(-k) \, 
v(p_1,p_2,k,k)\,v(k,k,\ell,\ell)\,.
\eea 
Using the  identity (see Appendix), 
\beq \label{appen3}
\,\int_{k}\,\tilde{\Delta}_R(k)\,\tilde{\Delta}_+(-k) \,
= -\frac12 \int_k\, \frac1{(k^2-m^2 +i\epsilon)^2}\,,
\eeq 
we may put this as 
\bea
&&
-i \Sigma_{(2a)}(p_1, p_2) = \frac{\lambda^2}4  \,
\Big(\frac49\Big)\, \int_{k,\ell} 
\frac i {(\ell^2 -m^2 +i\epsilon)(k^2 -m^2 +i\epsilon)^2} 
\cr 
&&\qquad \qquad \qquad\quad
 -\frac{\lambda^2}2 \,
\,\int_{k\,\ell}\tilde{\Delta}_+(\ell) \,\tilde{\Delta}_R(k)\,\tilde{\Delta}_+(-k) \,
 \Big(\frac19 \Big) \Bigg( 2 \cos
(k\wedge p_1 ) 
\cr 
&&\qquad\qquad \qquad \qquad\qquad\qquad 
+ 2 \cos (k\wedge \ell) +
\frac12 \cos (k\wedge (p_1 -\ell)) + \frac12 \cos(k\wedge(p+\ell)
\Bigg)\,. \qquad
\eea 
The first term is the planar diagram contribution
and is divergent with the factor reduced to 4/9. 
The divergence is absorbed in the mass 
and coupling constant renormalization. The
second term is the non-planar contribution and is again UV-IR
finite.

 The second contribution of the two loop is given as
 \beaq && -i
\Sigma_{(2b)}(p_1, p_2) = \frac1{3!}\quad
\begin{picture}(70,23)(-3,10)
\put(00,13){\line(3,0){60}}
\put(30,13){\circle{25}}
\put(-5,5){\footnotesize{$p_1$}}
\put(55,5){\footnotesize{$p_2$}}
\put(27,10.5){$ \triangleleft $}
\put(25,22){\footnotesize{$<$}}
\put(26,-1){\footnotesize{$<$}}
\end{picture}
+\frac1{3!}\quad
\begin{picture}(70,23)(-3,10)
\put(00,13){\line(3,0){60}}
\put(30,13){\circle{25}}
\put(-5,5){\footnotesize{$p_1$}}
\put(55,5){\footnotesize{$p_2$}}
\put(27,10.5){$\triangleright $}
\put(25.5,22){\footnotesize{$>$}}
\put(26,-1){\footnotesize{$>$}}
\end{picture}
\label{sigma2b}
\\ \nonumber \\ &&\qquad\qquad
= - \frac{\lambda^2}6\, \int_{k,\,\ell\,q} \,\, \tilde \Delta_+(k)
\tilde \Delta_R(\ell) \tilde \Delta_+(q)\, v(p_1, k, \ell, q)^2
\nonumber\\
&& \qquad\qquad  \qquad\qquad\qquad\qquad \times  \Bigg( (2\pi)^D
\delta(q+k+\ell+p_1) + p_1\leftrightarrow -p_1 \Bigg)\,. \nonumber
\eeaq Using the identity, \bea && \int_{k,\,\ell\,q} \,\, \tilde
\Delta_+(k) \tilde \Delta_R(\ell) \tilde \Delta_+(q)\, \, \Bigg(
(2\pi)^D \delta(q+k+\ell+p_1) + p_1\leftrightarrow  -p_1 \Bigg)
\\
&&\quad
= -i \int_{k,\,\ell\,q} \,\,
\frac 1{ (k^2-m^2+i\epsilon)
(\ell^2-m^2+i\epsilon)
(q^2-m^2+i\epsilon)}
(2\pi)^D \delta(q+k+\ell+p_1)
\eea
we may put (\ref{sigma2b}) as
\bea
&&-i \Sigma_{(2b)}(p_1, p_2) =
\frac{i\lambda^2}6\,
\Big(\frac16\Big)
\int_{k,\,\ell\,q} \,\,
\frac {(2\pi)^D \delta(q+k+\ell+p_1)}
{ (k^2-m^2+i\epsilon)
(\ell^2-m^2+i\epsilon)
(q^2-m^2+i\epsilon)}
\\ &&\qquad
- \frac{\lambda^2}6\,
\int_{k,\,\ell\,q} \,\,
\tilde \Delta_+(k)
\tilde \Delta_R(\ell)
\tilde \Delta_+(q)\,
\, \Bigg( (2\pi)^D \delta(q+k+\ell+p_1)
+ p_1\leftrightarrow  -p_1 \Bigg)\,
\\&&\qquad\qquad
\times \Big( \frac19 \Big)
\Bigg( \frac12 \cos(k+p)\wedge (p+q)
+   \frac12 \cos(k+p)\wedge (p+\ell)
+  \frac12  \cos(k+\ell)\wedge (k-q)
\\&&\qquad\qquad\qquad\qquad
+\cos \frac{(k+p)\wedge(p-k)}2
+\cos \frac{(k+p)\wedge(q-\ell)}2
+\cos \frac{(q+p)\wedge(p-q)}2
\\&&\qquad\qquad\qquad\qquad
+\cos \frac{(q+p)\wedge(k-\ell)}2 +\cos
\frac{(p+\ell)\wedge(p-\ell)}2 +\cos \frac{(p+\ell)\wedge(q-k)}2
\Bigg)\,. \eea The first term is the planar contribution and is
divergent while the rest is the non-planar contribution and
is UV-IR finite.

\subsection{Four point function}
\noindent

In this section, we provide a few diagramatic examples 
corresponding to the four point function.
The one loop correction to the four point function
is given as
\bea
-i\Gamma_4^{(1)} =
\frac12
\begin{picture} (100,40) (-10,25)
\put(04,10){\line(1,1){20}}
\put(04,50){\line(1,-1){20}}
\put(40,28){\circle{30}}
\put(56,28){\line(1,1){20}}
\put(56,28){\line(1,-1){20}}
\put(-8,12){\footnotesize{$p_1$}}
\put(-8,55){\footnotesize{$p_2$}}
\put(80,12){\footnotesize{$p_3$}}
\put(80,55){\footnotesize{$p_4$}}
\end{picture}
\; +\,\Big( p_2\leftrightarrow  p_3 \Big)
\,+ \, \Big(p_2\leftrightarrow  p_4 \Big)
\\ \eea
where each has 2-arrowed diagrams,
\bea
&& \begin{picture} (130,60) (-10,25)
\put(04,10){\line(1,1){20}}
\put(04,50){\line(1,-1){20}}
\put(40,28){\circle{30}}
\put(55.8,28){\line(1,1){20}}
\put(55.8,28){\line(1,-1){20}}
\put(-8,12){\footnotesize{$p_1$}}
\put(-8,55){\footnotesize{$p_2$}}
\put(80,12){\footnotesize{$p_3$}}
\put(80,55){\footnotesize{$p_4$}}
\put(38,50){\footnotesize{$k_1$}}
\put(38,0){\footnotesize{$k_2$}}
\end{picture}
=
\begin{picture} (130,60) (-10,25)
\put(04,10){\line(1,1){20}}
\put(04,50){\line(1,-1){20}}
\put(40,28){\circle{30}}
\put(36,41){$\triangleleft $}
\put(38,9.5){\footnotesize{$<$}}
\put(55.8,28){\line(1,1){20}}
\put(55.8,28){\line(1,-1){20}}
\put(-8,12){\footnotesize{$p_1$}}
\put(-8,55){\footnotesize{$p_2$}}
\put(80,12){\footnotesize{$p_3$}}
\put(80,55){\footnotesize{$p_4$}}
\put(38,50){\footnotesize{$k_1$}}
\put(38,0){\footnotesize{$k_2$}}
\end{picture}
+
\begin{picture} (130,60) (-10,25)
\put(05,10){\line(1,1){20}}
\put(05,50){\line(1,-1){20}}
\put(40,28){\circle{30}}
\put(36,41){$\triangleright $} \put(38,10){\footnotesize{$>$}}
\put(55,28){\line(1,1){20}}
\put(55,28){\line(1,-1){20}}
\put(-8,12){\footnotesize{$p_1$}}
\put(-8,55){\footnotesize{$p_2$}}
\put(80,12){\footnotesize{$p_3$}}
\put(80,55){\footnotesize{$p_4$}}
\put(38,50){\footnotesize{$k_1$}}
\put(38,0){\footnotesize{$k_2$}}
\end{picture}
\\ \\  \\
&&\quad= - \int_{k_1\,k_2}
\Bigg(\,v(p_1,p_2,k_1,k_2) \,\,
v(k_1,k_2,p_3,p_4) \,\,\tilde{\Delta}_R (k_1) \tilde{\Delta}_+ (k_2) \, (2\pi)^{D}
\delta^D (p_1 + p_2 + k_1 +k_2)  \,
\\&&\qquad\qquad\qquad
+ (p_1,p_2)\leftrightarrow  (p_3, p_4) \Bigg)\,. \eea

There are four types of two loop diagrams,
\beaq
\frac14
\begin{picture} (140,60) (-20,25)
\put(04,10){\line(1,1){20}}
\put(04,50){\line(1,-1){20}}
\put(40,28){\circle{30}}
\put(72,28){\circle{30}}
\put(88,28){\line(1,1){20}}
\put(88,28){\line(1,-1){20}}
\put(-8,12){\footnotesize{$p_1$}}
\put(-8,55){\footnotesize{$p_2$}}
\put(105,55){\footnotesize{$p_3$}}
\put(105,12){\footnotesize{$p_4$}}
\end{picture}
\; + \frac 14 \,
\begin{picture} (120,60) (-20,25)
\put(05,10){\line(1,1){20}}
\put(05,50){\line(1,-1){20}}
\put(41,30){\circle{30}}
\put(57,30){\line(1,1){20}}
\put(57,30){\line(1,-1){20}}
\put(41,56){\circle{20}}
\put(-8,12){\footnotesize{$p_1$}}
\put(-8,55){\footnotesize{$p_2$}}
\put(80,12){\footnotesize{$p_3$}}
\put(80,55){\footnotesize{$p_4$}}
\end{picture}
\cr
+
\frac 14 \,
\begin{picture}(120,60) (-20,25)
\put(-8,55){\footnotesize{$p_2$}}
\put(50,55){\footnotesize{$p_4$}}
\put(05,10){\line(1,1){40}}
\put(05,50){\line(1,-1){40}}
\put(42,29.5){\circle{25}}
\put(-8,8){\footnotesize{$p_1$}}
\put(50,8){\footnotesize{$p_3$}}
\end{picture}
+
\frac 14 \,
\begin{picture}(120,60) (-20,25)
\put(-8,55){\footnotesize{$p_2$}}
\put(50,55){\footnotesize{$p_4$}}
\put(05,10){\line(1,1){40}}
\put(05,50){\line(1,-1){40}}
\put(8,30){\circle{25}}
\put(-8,8){\footnotesize{$p_1$}}
\put(50,8){\footnotesize{$p_3$}}
\end{picture}
\cr\label{4-2loop}
\eeaq
and their crossed channels  $ p_2\leftrightarrow p_3 $ and
$p_2\leftrightarrow  p_4\,$.

The first diagram in (\ref{4-2loop})
has the 4 distinct arrowed diagrams,
whose contributions
are given as
\bea
&&
\begin{picture} (140,30) (-10,30)
\put(06,10){\line(1,1){20}}
\put(06,50){\line(1,-1){20}}
\put(42,28){\circle{30}}
\put(74,28){\circle{30}}
\put(90,28){\line(1,1){20}}
\put(90,28){\line(1,-1){20}}
\put(-8,12){\footnotesize{$p_1$}}
\put(-8,55){\footnotesize{$p_2$}}
\put(115,55){\footnotesize{$p_3$}}
\put(115,10){\footnotesize{$p_4$}}
\put(40,50){\footnotesize{$k_1$}}
\put(40,3){\footnotesize{$k_2$}}
\put(75,50){\footnotesize{$\ell_1$}}
\put(75,3){\footnotesize{$\ell_2$}}
\end{picture}
\\ \\ \\
&&=
\begin{picture} (140,30) (-10,30)
\put(06,10){\line(1,1){20}}
\put(06,50){\line(1,-1){20}}
\put(42,28){\circle{30}}
\put(74,28){\circle{30}}
\put(90,28){\line(1,1){20}}
\put(90,28){\line(1,-1){20}}
\put(-8,12){\footnotesize{$p_1$}}
\put(-8,55){\footnotesize{$p_2$}}
\put(115,55){\footnotesize{$p_3$}}
\put(115,10){\footnotesize{$p_4$}}
\put(40,50){\footnotesize{$k_1$}}
\put(40,0){\footnotesize{$k_2$}}
\put(75,50){\footnotesize{$\ell_1$}}
\put(75,0){\footnotesize{$\ell_2$}}
\put(38.5,41){$\triangleleft $}
\put(40,10){\footnotesize{$<$}}
\put(71,41){$\triangleleft $}
\put(70,10){\footnotesize{$<$}}
\end{picture}
+
\begin{picture} (140,30) (-10,30)
\put(06,10){\line(1,1){20}}
\put(06,50){\line(1,-1){20}}
\put(42,28){\circle{30}}
\put(74,28){\circle{30}}
\put(90,28){\line(1,1){20}}
\put(90,28){\line(1,-1){20}}
\put(-8,12){\footnotesize{$p_1$}}
\put(-8,55){\footnotesize{$p_2$}}
\put(115,55){\footnotesize{$p_3$}}
\put(115,10){\footnotesize{$p_4$}}
\put(40,50){\footnotesize{$k_1$}}
\put(40,0){\footnotesize{$k_2$}}
\put(75,50){\footnotesize{$\ell_1$}}
\put(75,0){\footnotesize{$\ell_2$}}
\put(39,41){$\triangleright $}
\put(40,10){\footnotesize{$>$}}
\put(71,41){$\triangleright $}
\put(70,10){\footnotesize{$>$}}
\end{picture}
\\ \\ \\
&& \qquad\qquad+
\begin{picture} (140,30) (-10,30)
\put(06,10){\line(1,1){20}}
\put(06,50){\line(1,-1){20}}
\put(42,28){\circle{30}}
\put(74,28){\circle{30}}
\put(90,28){\line(1,1){20}}
\put(90,28){\line(1,-1){20}}
\put(-8,12){\footnotesize{$p_1$}}
\put(-8,55){\footnotesize{$p_2$}}
\put(115,55){\footnotesize{$p_3$}}
\put(115,10){\footnotesize{$p_4$}}
\put(40,50){\footnotesize{$k_1$}}
\put(40,0){\footnotesize{$k_2$}}
\put(75,50){\footnotesize{$\ell_1$}}
\put(75,0){\footnotesize{$\ell_2$}}
\put(38,41){$\triangleleft $}
\put(40,9.5){\footnotesize{$<$}}
\put(69,41){$\triangleright $}
\put(69,9.8){\footnotesize{$>$}}
\end{picture}
+
\begin{picture} (140,30) (-10,30)
\put(06,10){\line(1,1){20}}
\put(06,50){\line(1,-1){20}}
\put(42,28){\circle{30}}
\put(74,28){\circle{30}}
\put(90,28){\line(1,1){20}}
\put(90,28){\line(1,-1){20}}
\put(-8,12){\footnotesize{$p_1$}}
\put(-8,55){\footnotesize{$p_2$}}
\put(115,55){\footnotesize{$p_3$}}
\put(115,10){\footnotesize{$p_4$}}
\put(40,50){\footnotesize{$k_1$}}
\put(40,0){\footnotesize{$k_2$}}
\put(75,50){\footnotesize{$\ell_1$}}
\put(75,0){\footnotesize{$\ell_2$}}
\put(38,41){$\triangleright $}
\put(40,10){\footnotesize{$>$}}
\put(71,41){$\triangleleft $}
\put(70,10){\footnotesize{$<$}}
\end{picture}
\\ \\ \\
&&
=-i \int_{k_1,k_2,\ell_1,\ell_2}
v(p_1,p_2,k_1,k_2)
v(k_1,k_2,\ell_1,\ell_2)
v(\ell_1, \ell_2, p_3, p_4)
\, \tilde{\Delta}_R (k_1) \tilde{\Delta}_R (\ell_1)
\tilde{\Delta}_+(k_2) \tilde{\Delta}_+(\ell_2)
\\
\\&&\qquad \times
(2\pi)^{2D} \Big(  \delta^D(p+k) +  \delta^D(p-k)\Big) \Big(
\delta^D(k-\ell) + \delta^D(k+\ell) \Big) 
\eea 
where $p=p_1+p_2
=p_3+ p_4 $, $k=k_1+k_2$ and $\ell=\ell_1 + \ell_2$.

The second diagram in (\ref{4-2loop})
also has the 6 distinct diagrams,
\bea
&&\begin{picture} (130,50) (-10,30)
\put(05,10){\line(1,1){20}}
\put(05,50){\line(1,-1){20}}
\put(42,28){\oval(34,25)}
\put(41,51){\circle{20}}
\put(59,28){\line(1,1){20}}
\put(59,28){\line(1,-1){20}}
\put(-8,12){\footnotesize{$p_1$}}
\put(-8,55){\footnotesize{$p_2$}}
\put(80,12){\footnotesize{$p_3$}}
\put(80,55){\footnotesize{$p_4$}}
\put(18,45){\footnotesize{$k_1$}}
\put(35,5){\footnotesize{$k_2$}}
\put(35,65){\footnotesize{$\ell$}}
\put(55,45){\footnotesize{$k_1$}}
\end{picture}\\ \\ \\
&& =
\begin{picture} (130,50) (-10,30)
\put(05,10){\line(1,1){20}}
\put(05,50){\line(1,-1){20}}
\put(59,28){\line(1,1){20}}
\put(59,28){\line(1,-1){20}}
\put(42,28){\oval(34,25)}
\put(41,51){\circle{20}}
\put(-8,10){\footnotesize{$p_1$}}
\put(-8,55){\footnotesize{$p_2$}}
\put(80,10){\footnotesize{$p_3$}}
\put(80,55){\footnotesize{$p_4$}}
\put(18,45){\footnotesize{$k_1$}}
\put(27,3){\footnotesize{$k_2$}}
\put(35,65){\footnotesize{$\ell$}}
\put(55,45){\footnotesize{$k_1$}}
\put(37,58.5){\footnotesize{$>$}}
\put(27,35.8){\footnotesize{$\triangle$}}
\put(51,35){\footnotesize{$\bigtriangledown$}}
\put(38,13.3){\footnotesize{$<$}}
\end{picture}
+
\begin{picture} (130,50) (-10,30)
\put(05,10){\line(1,1){20}}
\put(05,50){\line(1,-1){20}}
\put(59,28){\line(1,1){20}}
\put(59,28){\line(1,-1){20}}
\put(42,28){\oval(34,25)}
\put(41,51){\circle{20}}
\put(-8,10){\footnotesize{$p_1$}}
\put(-8,55){\footnotesize{$p_2$}}
\put(80,10){\footnotesize{$p_3$}}
\put(80,55){\footnotesize{$p_4$}}
\put(18,45){\footnotesize{$k_1$}}
\put(27,3){\footnotesize{$k_2$}}
\put(35,65){\footnotesize{$\ell$}}
\put(55,45){\footnotesize{$k_1$}}
\put(35,59){\footnotesize{$>$}}
\put(24.5,35){\footnotesize{$\bigtriangledown$}}
\put(49,36){\footnotesize{$\triangle$}}
\put(38,13.5){\footnotesize{$>$}}
\end{picture}
\\ \\ \\
&& \qquad +
\begin{picture} (130,50) (-10,30)
\put(05,10){\line(1,1){20}}
\put(05,50){\line(1,-1){20}}
\put(59,28){\line(1,1){20}}
\put(59,28){\line(1,-1){20}}
\put(42,28){\oval(34,25)}
\put(41,51){\circle{20}}
\put(-8,10){\footnotesize{$p_1$}}
\put(-8,55){\footnotesize{$p_2$}}
\put(80,10){\footnotesize{$p_3$}}
\put(80,55){\footnotesize{$p_4$}}
\put(18,45){\footnotesize{$k_1$}}
\put(27,3){\footnotesize{$k_2$}}
\put(35,65){\footnotesize{$\ell$}}
\put(55,45){\footnotesize{$k_1$}}
\put(36.5,58.6){\footnotesize{$>$}}
\put(25.8,33.5){\line(0,1){5}}
\put(25.8,33.5){\line(1,0){4}}
\put(50,35){\footnotesize{$\bigtriangledown$}}
\put(39,13) {$\triangleleft$}
\end{picture}
+
\begin{picture} (130,50) (-10,30)
\put(05,10){\line(1,1){20}}
\put(05,50){\line(1,-1){20}}
\put(59,28){\line(1,1){20}}
\put(59,28){\line(1,-1){20}}
\put(42,28){\oval(34,25)}
\put(41,51){\circle{20}}
\put(-8,10){\footnotesize{$p_1$}}
\put(-8,55){\footnotesize{$p_2$}}
\put(80,10){\footnotesize{$p_3$}}
\put(80,55){\footnotesize{$p_4$}}
\put(18,45){\footnotesize{$k_1$}}
\put(27,3){\footnotesize{$k_2$}}
\put(35,65){\footnotesize{$\ell$}}
\put(55,45){\footnotesize{$k_1$}}
\put(36.5,58.5){\footnotesize{$>$}}
\put(24.5,34.5){\footnotesize{$\bigtriangledown$}}
\put(58,34){\line(0,1){4}}
\put(54,33){\line(1,0){4}}
\put(40,13){\footnotesize{$\triangleright$}}
\end{picture}
\\ \\ \\
&& \qquad +
\begin{picture} (130,50) (-10,30)
\put(05,10){\line(1,1){20}}
\put(05,50){\line(1,-1){20}}
\put(59,28){\line(1,1){20}}
\put(59,28){\line(1,-1){20}}
\put(42,28){\oval(34,25)}
\put(41,51){\circle{20}}
\put(-8,10){\footnotesize{$p_1$}}
\put(-8,55){\footnotesize{$p_2$}}
\put(80,10){\footnotesize{$p_3$}}
\put(80,55){\footnotesize{$p_4$}}
\put(18,45){\footnotesize{$k_1$}}
\put(27,3){\footnotesize{$k_2$}}
\put(35,65){\footnotesize{$\ell$}}
\put(55,45){\footnotesize{$k_1$}}
\put(37,58.5){\footnotesize{$>$}}
\put(27,36){\footnotesize{$\triangle$}}
\put(54,39){\line(1,0){4}}
\put(54,34){\line(0,1){5}}
\put(40,12.5) {$\triangleleft$}
\end{picture}
+
\begin{picture} (130,50) (-10,30)
\put(05,10){\line(1,1){20}}
\put(05,50){\line(1,-1){20}}
\put(59,28){\line(1,1){20}}
\put(59,28){\line(1,-1){20}}
\put(42,28){\oval(34,25)}
\put(41,51){\circle{20}}
\put(-8,10){\footnotesize{$p_1$}}
\put(-8,55){\footnotesize{$p_2$}}
\put(80,10){\footnotesize{$p_3$}}
\put(80,55){\footnotesize{$p_4$}}
\put(18,45){\footnotesize{$k_1$}}
\put(27,3){\footnotesize{$k_2$}}
\put(35,65){\footnotesize{$\ell$}}
\put(55,45){\footnotesize{$k_1$}}
\put(37,58.5){\footnotesize{$>$}}
\put(26,38.5){\line(1,0){4}}
\put(30.5,34){\line(0,1){4}}
\put(48,37){\footnotesize{$\triangle$}}
\put(39,13) {$\triangleright$}
\end{picture}
\\ \\ \\
&&= -i \int_{k_1,k_3, k_4}
v(p_1,p_2,k_1,k_2)
v(k_1,k_1,\ell,\ell)
v(k_1, k_2, p_3, p_4)\,\tilde{\Delta}_+(\ell)  \tilde{\Delta}_R(k_1)\,
\\
\\&&\qquad\qquad
\times (2\pi)^{D}
\Bigg( \Big(\tilde{\Delta}_R(k_1) \tilde{\Delta}_+(k_2)
+ \tilde{\Delta}_+ (k_1)  \tilde{\Delta}_R(k_2) \Big) \,
\Big( \delta^D (p + k) +  \delta^D (p - k) \Big)
\\&& \qquad\qquad\qquad\qquad
+ \tilde{\Delta}_+ (k_1)  \tilde{\Delta}_R(k_2) \,
(\delta^D (p -k_1+ k_2) +  \delta^D (p -k_1- k_2)  \Bigg)
\,.
\eea

The third diagram in (\ref{4-2loop})
has 6 distinct diagrams,
\bea
&&
\begin{picture}(120,60) (30,25)
\put(-8,45){\footnotesize{$p_2$}}
\put(90,55){\footnotesize{$p_4$}}
\put(05,10){\line(2, 1){80}}
\put(05,40){\line(2,-1){80}}
\put(71.5,25){\circle{30}}
\put(42,39){\footnotesize{$k_1$}}
\put(42,5){\footnotesize{$k_2$}}
\put(62,24){\footnotesize{$\ell_1$}}
\put(92,24){\footnotesize{$\ell_2$}}
\put(-8,8){\footnotesize{$p_1$}}
\put(90,0){\footnotesize{$p_3$}}
\end{picture}
\\ \\
&& =
\begin{picture}(120,60) (-20,25)
\put(-8,45){\footnotesize{$p_2$}}
\put(90,55){\footnotesize{$p_4$}}
\put(05,10){\line(2, 1){80}}
\put(05,40){\line(2,-1){80}}
\put(72,25.5){\circle{30}}
\put(-8,8){\footnotesize{$p_1$}}
\put(90,0){\footnotesize{$p_3$}}
\put(42,39){\footnotesize{$k_1$}}
\put(42,5){\footnotesize{$k_2$}}
\put(62,24){\footnotesize{$\ell_1$}}
\put(92,24){\footnotesize{$\ell_2$}}
\put(43,21){\line(1,0){5}}
\put(43,17){\line(0,1){4}}
\put(52,23){\footnotesize{$\triangle$}}
\put(84.5,24){\footnotesize{$\vee$}}
\put(42,28){\footnotesize{$\triangle$}}
\end{picture}
+
\begin{picture}(120,60) (-20,25)
\put(-8,45){\footnotesize{$p_2$}}
\put(90,55){\footnotesize{$p_4$}}
\put(05,10){\line(2, 1){80}}
\put(05,40){\line(2,-1){80}}
\put(72,25.5){\circle{30}}
\put(-8,8){\footnotesize{$p_1$}}
\put(90,0){\footnotesize{$p_3$}}
\put(42,39){\footnotesize{$k_1$}}
\put(42,5){\footnotesize{$k_2$}}
\put(62,24){\footnotesize{$\ell_1$}}
\put(92,24){\footnotesize{$\ell_2$}}
\put(53,24){\footnotesize{$\vee$}}
\put(85,24){\footnotesize{$\vee$}}
\put(42,28){\footnotesize{$\triangle$}}
\put(42,17){\footnotesize{$\bigtriangledown$}}
\end{picture}
\\ \\
&& \qquad +
\begin{picture}(120,60) (-20,25)
\put(-8,45){\footnotesize{$p_2$}}
\put(90,55){\footnotesize{$p_4$}}
\put(05,10){\line(2, 1){80}}
\put(05,40){\line(2,-1){80}}
\put(72,25.5){\circle{30}}
\put(-8,8){\footnotesize{$p_1$}}
\put(90,0){\footnotesize{$p_3$}}
\put(42,39){\footnotesize{$k_1$}}
\put(42,5){\footnotesize{$k_2$}}
\put(62,24){\footnotesize{$\ell_1$}}
\put(92,24){\footnotesize{$\ell_2$}}
\put(42,18){\line(1,0){5}}
\put(47,18.5){\line(0,1){5}}
\put(42,28.6){\footnotesize{$\triangle$}}
\put(51.5,23.5){\footnotesize{$\bigtriangledown$}}
\put(85,24){\footnotesize{$\vee$}}
\end{picture}
\qquad + \qquad  (p_3 \leftrightarrow p_4)
 \\ \\ \\
&& = -i \int_{k_1,k_2,k_3, k_4}
v(p_1,p_2,k_1,k_2)
v(k_1,\ell_1,\ell_2,p_4)
v(k_2, \ell_1, \ell_2, p_3)\,
\\
&&\qquad \times (2\pi)^{2D} \,
\tilde{\Delta}_+(\ell_2)\,
\Big( \tilde{\Delta}_R(k_1)\tilde{\Delta}_+(k_2)
 \tilde{\Delta}_R(\ell_1)
\delta^D(p+k) \delta^D(p_3 -k_1 + \ell)
\\&& \qquad\qquad\qquad\qquad\quad
+ \tilde{\Delta}_R(k_1) \tilde{\Delta}_R(k_2)
\tilde{\Delta}_+(\ell_1)
\delta^D(p-k_1+k_2) \delta^D(p_3 +k_1 + \ell)
\\ &&\qquad\qquad\qquad\qquad\quad
+ \tilde{\Delta}_R(k_1) \tilde{\Delta}_+(k_2)
\tilde{\Delta}_R(\ell_1)
\delta^D(p-k) \delta^Dq
(p_3 +k_1 - \ell)  \Big)
\\&& \qquad
+ (p_3\leftrightarrow  p_4)\,.
\eea

The fourth diagram in (\ref{4-2loop})
is the same as the third diagram
with $(p_1,p_2) \leftrightarrow (p_3, p_4)$.

\vskip.5cm
\section{Conclusion and outlook\label{c}}

The unitary S-matrix has been constructed in space-time non-commutative
field theory by introducing a proper treatment of the
time-ordering, the so-called minimal realization of the
time-ordering and $\star$-time ordering. 
Based on this unitary
S-matrix, the Feynman rule is established for the perturbation of
STNC real scalar field theory.
We note that our time-ordering differs from the 
one suggested or conjectured in \cite{bahns,topt};
their S-matrix can be unitary but 
will not guarantee the Heisenberg equation of motion
as requested in the Yang-Feldmann approach \cite{yf}.

Loop calculations of the STNC theory demonstrate that the divergent
structure is the same as in the SSNC theory, which comes from the
planar diagrams. The non-planar diagrams are finite as in the SSNC
real scalar field theory and remarkably, there is no UV/IR mixing
problem in the STNC result.

The perturbation theory is not limited to the real scalar theory.
One may generalize this formalism to complex scalar field theory,
fermionic theory, and gauge theory. Especially, the gauge theory
possesses derivative interaction and needs further care such as in
time-ordering and gauge symmetry. The details of which are in
preparation and will be published elsewhere \cite{ncgauge}.

Finally, it is noted that the formalism is considered so far in
terms of the Lagrangian formalism of the second quantized
operators in the Heisenberg picture. 
The Hamiltonian formalism is not easy to obtain \cite{ham}
and there lacks the path-integral formalism.  
The path integral formalism is necessary to
accommodate the non-abelian gauge theory. Currently, finding the
path-integral approach of the theory looks a very challenging
problem to solve.

\vskip 1cm

{\sl Acknowledgement:}
It is acknowledged that this work was supported in part
by the Basic Research Program of the Korea Science
and Engineering Foundation Grant number
R01-1999-000-00018-0(2003)(CR),
R01-2003-000-10391-0 (YS),
and by Korea Research Foundation
under project number KRF-2003-005-C00010 (JHY).
CR is also grateful for KIAS
while this work is done during his visit.

\section*{Appendix}
In this appendix, we provide some identities 
useful for loop correlations. \par
\vskip 0.3cm
{\sl 1. Identity of Eq.~(\ref{appen1})}: 
The left hand side of (\ref{appen1}) becomes
\bea
LHS &=& \int_{k} \tilde{\Delta}_{+}(k) 
= \int {d^D k \over (2\pi)^D} 2\pi 
\,\delta(k^2 -m^2)\theta(k_{0}) \cr
&=& \int {d^D k \over (2\pi)^D} 2\pi \,
\delta\left(k_{0}^{2} -\omega_{k}^2\right)\theta(k_{0})\cr
&=& \int {d^{D-1} \vec{k} \over (2\pi)^{D-1}} {1 \over 2 \omega_{k}}, 
\eea
where we use the following identity
\bea
\delta(y^2 -a^2) =|2a|^{-1} \left[\delta(y-a)+\delta(y+a)\right].
\eea
The right hand side of (\ref{appen1}) is given as;
\bea
RHS = 
i \int_k \frac{1}{k^2 -m^2 + i\epsilon}
=  i 
\int \frac{d^{D-1} \vec{k}}{(2 \pi)^{D-1}} \, 
\int \frac{d k_{0}}{2\pi}\,\,
\frac{1}{ (k_{0} -\omega_{k} +i\delta) 
( k_{0} +\omega_{k} - i\delta)}\,.
\eea
To evaluate this, one may use contour integral over $k_0$.
The contour integral on the upper half-plane  
has the contribution from the positive imaginary pole:
\bea
RHS =  \int \frac{d^{D-1} \vec{k}}{(2 \pi)^{D-1}} \, 
\frac{1}{2 \omega_{k}}\,.
\eea
Hence, (\ref{appen1})  follows:
\bea
\int_{k} \tilde{\Delta}_{+} 
= i \int_{k} \frac{1}{k^2 -m^2 +i \epsilon}.
\eea

\vskip 0.3cm
{\sl 2. Identity of (\ref{appen2})}: 
We can derive this identity by taking advantage of  
the delta function in $\tilde{\Delta}_{+}(k)$. 
\bea
LHS &=& \int_{k} \tilde{\Delta}_{+}(k) \cos(p\wedge k) 
= \int_{k} 2\pi \delta(k^2 -m^2) \theta(k_{0}) 
\cos(p_{0} k^{(N)} \theta -p^{(N)} k_{0} \theta)
\cr 
&=&  \int_{k} \theta(k_{0}) 2\pi \delta(k^2 -m^2) 
\cos(p_{0} k^{(N)}  \theta) \, \cos(p^{(N)}  k_{0} \theta)
\cr
&=& \frac 12  \int_{k} 2\pi \delta(k^2 -m^2) 
\cos(p_{0} k^{(N)}  \theta) \, \cos(p^{(N)}  k_{0} \theta)
\eea
where we use the symmetry 
of spatial component of $k$:
${\bf k} \to -{\bf k}$
in the second line.
$p^{(N)} $ or $ k^{(N)} $  
refers to the non-commuting component of the 
spatial momentum.
Using the integral representation 
of the delta-function we have
\bea
LHS &=&
\frac 12  \int_{k} 
\cos(p_{0} k^{(N)}  \theta) \, \cos(p^{(N)}  k_{0} \theta)
\int_{-\infty}^{\infty} d\alpha\, e^{i \alpha (k^2 -m^2)} \cr
&=& \frac 12  
\int_{k} \cos(p_{0} k^{(N)}  \theta) \, \cos(p^{(N)}  k_{0} \theta)
\,\int_{0}^{\infty} d\alpha\,  e^{i \alpha (k^2 -m^2)}  + c.c. 
\cr
&=& \frac{1}{2} \,\int_{0}^{\infty} d\alpha\,  
e^{-i\alpha m^2}\,
\int_{k_0} \, e^{i\alpha k_0^2 }\, \cos(p^{(N)}  k_{0} \theta)
\int_{\bf k} \, e^{-i\alpha {\bf k}^2 }\,\cos(p_{0} k^{(N)}  \theta)\,.
\eea
$k_0$ integration becomes
\bea
\int \frac{d k_{0}}{2\pi}\, 
e^{i\alpha k_0^2 }\, \cos(p^{(N)}  k_{0} \theta)
=  e^{ - i \frac{(p^{(N)}  \, \theta)^2 }{4 \alpha} }
\int \frac{d k_{0}}{2\pi}\,
e^{i\alpha k_0^2 }
=  e^{ - i \frac{(p^{(N)}  \, \theta)^2 }{4 \alpha} }\,
\frac{e^{i \pi/4}}{\sqrt{4 \pi \alpha}} 
\eea
where we shift $k_{0}$
by $k_{0} \pm p^{(N)} \theta/(2\alpha)$ 
in the first identity, and rotate 
$ k_{0}$ by 
$\frac{e^{i\pi/ 4}}{\sqrt{\alpha}} \,k_{0}$
in the last identity to evaluate the Gaussian integral.
Likewise, we have
\bea
\int \frac{d^{D-1} \vec{k}}{(2 \pi)^{D-1}} \, 
e^{-i\alpha {\bf k}^2 }\,\cos(p_{0} k^{(N)} \theta)\,
= e^{ i \frac{(p_0 \, \theta)^2 }{4 \alpha} }
\int \frac{d^{D-1} \vec{k}}{(2 \pi)^{D-1}} \, 
e^{- i\alpha {\bf k}^2 }
=e^{ i \frac{(p_0 \, \theta)^2 }{4 \alpha} }\,
\Bigg(\frac{e^{-i \pi/4}}{\sqrt{4 \pi \alpha}} \Bigg)^{D-1}
\eea
after shifting 
 $k^{(N)} $ by 
$k^{(N)} \pm  p_{0}\theta/(2\alpha)$, 
and rotating ${\bf k}$
by $\frac{e^{-i\pi/4}} {\sqrt{\alpha}} {\bf k}$.
Hence, we have 
\bea
LHS = \frac{1}{2} \int_{0}^{\infty} \,
d \alpha \,\,\,
\frac{e^{-i\pi(D-2) /4 }} {(4\pi \alpha)^{D/2}} \,\,\,
e^{-i \alpha m^2 + \frac{i\,p\circ p}{4\alpha}}\,
+ c.c.
\eea
where $p\circ p \equiv (p_{0} \theta)^2 -(\vec{p} \theta)^2$. 
Finally, one more rotation of $\alpha$ 
to $ e^{-\frac{i\pi}{2}}\alpha$
gives the desired result;
\bea
LHS =\frac{1}{2} \int_{0}^{\infty} \,d\alpha \,
\frac{e^{-\alpha m^2 -\frac{p\circ p}{4\alpha}}}{(4\pi\alpha)^{D/2}}
+c.c.
= \int_{0}^{\infty} 
\, \frac{d\alpha} {(4\pi\alpha)^{D/2}}\,\,\,
e^{-\alpha m^2 -\frac{p\circ p}{4\alpha}}\,.
\eea
Therefore, we have 
\bea
\int_{k} \tilde{\Delta}_{+}(k)\cos(p\wedge k)=
\int_{0}^{\infty} 
\, \frac{d\alpha} {(4\pi\alpha)^{D/2}}\,\,\,
e^{-\alpha m^2 -\frac{p\circ p}{4\alpha}}\,.
\eea

\vskip 0.3cm
{\sl 3. Identity of (\ref{appen3})}: 
Left hand side is integrated over $k_0$  
using the delta-function:
\bea
LHS &=& 
\int_{k} \tilde{\Delta}_{+}(-k)\tilde{\Delta}_{R}(k)
\cr &=&  \int {d^D k \over (2\pi)^D} \, 
2\pi \delta(k^2 -m^2)\theta(-k_{0}) \,
\frac{i}{2\omega_{k}}\frac{1}{k_{0}-\omega_{k}+i\epsilon} \cr
&=& - i \int {d^{D-1} k \over (2\pi)^{D-1}} 
\frac{1}{(2\omega_{k})^3}. 
\eea
One integrates over $k_0$ 
in the right hand side by the contour integral:
\bea
\int_{k} \frac{1}{(k^2 -m^2 +i\epsilon)^2} 
&=& i \int \frac{d^{D-1} k}{(2 \pi)^{D-1}} \,
\frac{d}{dk_{0}} \Big(\frac{1}{k_{0} -\omega_{k} +i\epsilon}\Big)^2 
\Bigg|_{k_0=-\omega_k} \cr
&=& 2 i \int {d^{D-1} \vec{k} \over (2\pi)^{D-1}} \frac{1}{(2\omega_{k})^3}. 
\eea
Hence the identity follows: 
\bea
\int_{k}\tilde{\Delta}_{+}(-k) \tilde{\Delta}_{R}(k) = -\frac{1}{2}\int_{k} \frac{1}{(k^2 -m^2 +i\epsilon)^2}.
\eea

\end{document}